\documentclass[sigplan,10pt]{acmart}
\usepackage{xcolor}
\usepackage{xspace}
\usepackage{mathtools} 

\usepackage{amssymb}
\usepackage{bm}
\usepackage{enumitem}
\usepackage{balance}
\usepackage{textcomp}
\usepackage[]{hyperref}

\DeclareMathOperator*{\argmin}{arg\,min}

\definecolor{darkgreen}{HTML}{336633}

\usepackage{graphicx}
\usepackage{subcaption}
\usepackage{tikz}
\usepackage[labelfont=bf]{caption}

\usepackage{tabularx}
\usepackage{multirow}
\usepackage{booktabs}
\usepackage{makecell}
\usepackage[figuresleft]{rotating}
\usepackage[flushleft]{threeparttable}
\usepackage{colortbl}

\usepackage{algorithmicx}
\usepackage{algorithm}
\usepackage[noend]{algpseudocode}

\algnewcommand\algorithmicinput{\textbf{Input:}}
\algnewcommand\Input{\item[\algorithmicinput]}
\algnewcommand\algorithmicoutput{\textbf{Output:}}
\algnewcommand\Output{\item[\algorithmicoutput]}

\usepackage[utf8]{inputenc} 
\usepackage[most]{tcolorbox} 
\usepackage{soul} 
\usepackage{siunitx} 
\usepackage{pifont} 
\usepackage{xurl} 

\usepackage[noindentafter]{titlesec}
\titlespacing\section{0pt}{3pt}{3pt} 
\titlespacing\subsection{0pt}{3pt}{3pt}
\titlespacing\subsubsection{0pt}{5pt}{5pt}

\usepackage[skip=3pt]{caption}
\lstdefinestyle{code}{
    belowcaptionskip=1\baselineskip,
    breaklines=true,
    frame=none,
    numbers=none, 
    basicstyle=\footnotesize\ttfamily,
    keywordstyle=\bfseries\color{green!40!black},
    commentstyle=\itshape\color{purple!40!black},
    identifierstyle=\color{black},
    backgroundcolor=\color{white},
}
\setcopyright{rightsretained}
\begin{document}
\date{}
\newcommand{\circledtext}[1]{\textcircled{\raisebox{-0.9pt}{#1}}}

\author{Xiaozhe Yao}
\email{xiaozhe.yao@inf.ethz.ch}
\orcid{0000-0002-4661-533X}
\affiliation{\institution{ETH Zurich}\country{Switzerland}}

\author{Qinghao Hu}
\email{qinghao@mit.edu}
\orcid{0000-0003-1034-7502}
\affiliation{\institution{MIT}\country{United States}}

\author{Ana Klimovic}
\email{aklimovic@ethz.ch}
\orcid{0000-0001-8559-0529}
\affiliation{\institution{ETH Zurich}\country{Switzerland}}

\newcommand{\rawprojectname}{\textsc{DeltaZip}}
\newcommand{\projectname}{\rawprojectname\xspace}
\newcommand{\rawalgname}{$\Delta$\textsc{Compress}}
\newcommand{\algname}{$\Delta$\textsc{Compress}\xspace}
\newcommand{\papername}{\projectname: Efficient Serving of Multiple Full-Model-Tuned LLMs}

\acmYear{2025}\copyrightyear{2025}

\acmConference[EuroSys '25]{Twentieth European Conference on Computer Systems}{March 30--April 3, 2025}{Rotterdam, Netherlands}
\acmBooktitle{Twentieth European Conference on Computer Systems (EuroSys '25), March 30--April 3, 2025, Rotterdam, Netherlands}
\acmDOI{10.1145/3689031.3717468}
\acmISBN{979-8-4007-1196-1/25/03}

\title{\papername}
\begin{abstract}
    Fine-tuning large language models (LLMs) greatly improves model quality for downstream tasks. However, serving many fine-tuned LLMs concurrently is challenging due to the sporadic, bursty, and varying request patterns of different LLMs. To bridge this gap, we present \projectname, an LLM serving system that efficiently serves multiple full-parameter fine-tuned models concurrently by aggressively compressing model deltas by up to 10$\times$ while maintaining high model quality. The key insight behind this design is that fine-tuning results in small-magnitude changes to the pre-trained model. By co-designing the serving system with the compression algorithm,  \projectname achieves 2$\times$ to 12$\times$ improvement in throughput compared to the state-of-the-art systems. 

\end{abstract}



\maketitle

\pagestyle{plain}

\section{Introduction}
\label{sec:introduction}

Large Language Models (LLMs), such as GPT~\cite{openai_gpt-4_2023}, Llama~\cite{touvron_llama_2023} and Gemini~\cite{gemini_2024} have demonstrated remarkable performance and are widely used in a variety of applications, such as chatbots~\cite{openai_introducing_nodate} and coding assistants~\cite{github_github_nodate, chen_evaluating_2021, zhou_llm_2023}. To achieve high accuracy for a target domain, LLMs are first pre-trained on a large corpus of text data, then fine-tuned on application-specific tasks or datasets~\cite{liu_goat_2023, raffel_exploring_2023}, such as code~\cite{chen_evaluating_2021}, conversations~\cite{ouyang_training_2022}, and human preferences~\cite{rafailov_direct_2023}.
Cloud AI infrastructure companies, such as OpenAI~\cite{openai_gpt-35_nodate}, Google~\cite{google_cloud_platform_vertex_nodate}, Microsoft~\cite{madiepev_fine-tune_nodate}, and Anyscale~\cite{anyscale_fine-tuning_2023} expose APIs for users to fine-tune a pre-trained LLM with their own data and deploy the resulting customized model variant for inference.

\begin{figure}[tp]
  \includegraphics[width=\linewidth,trim={6cm 6cm 1cm 2cm},clip]{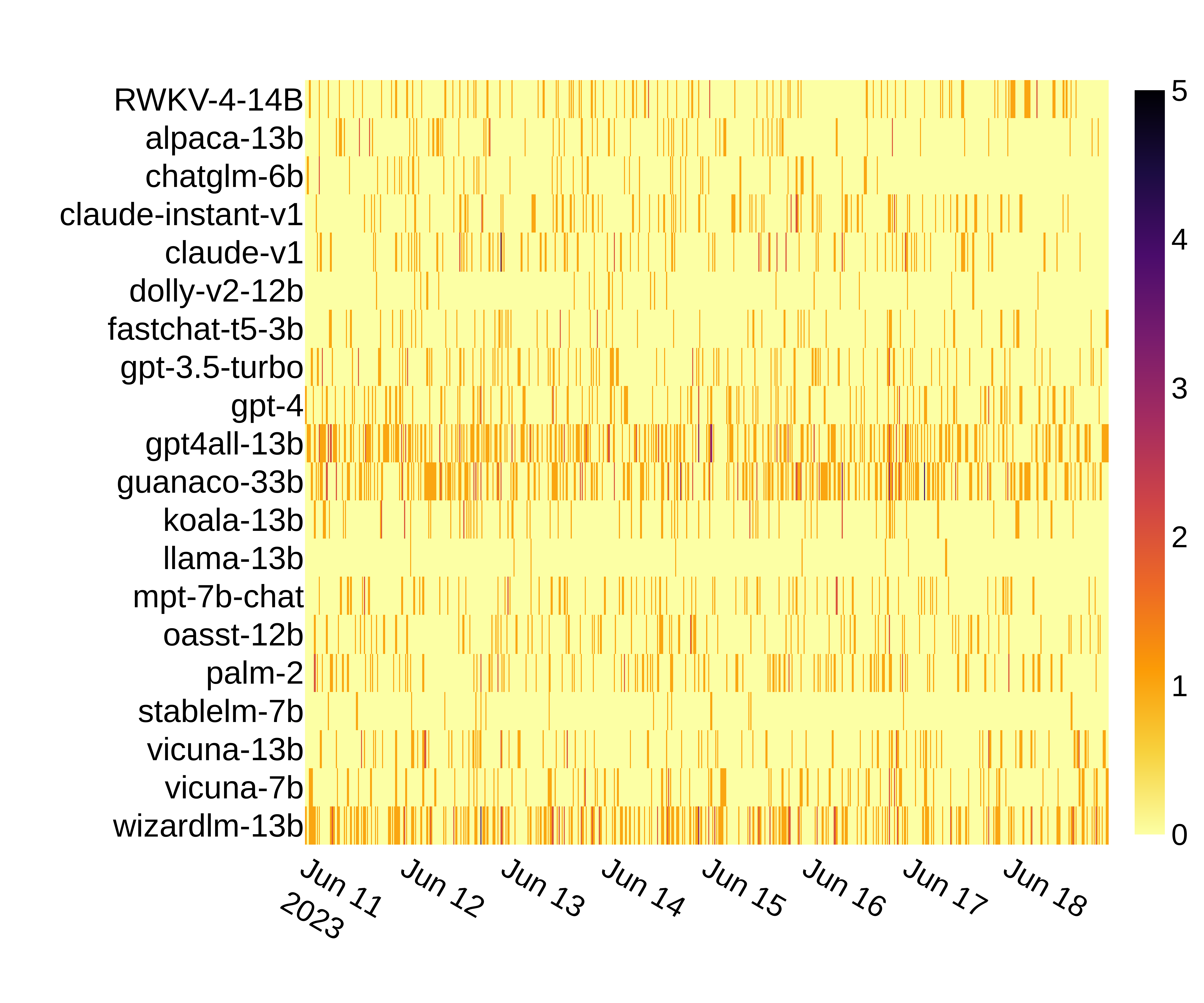}
  \caption{Invocation counts per 5-min time windows for 20 different models in the LMSys Chatbot Arena~\cite{zheng_judging_2023} trace.}
  \label{fig:motivation_trace}
\end{figure}

While fine-tuning is typically a one-time effort performed off the critical path, LLM serving is critical to optimize as it is typically recurring and latency-critical. Techniques such as continuous batching~\cite{yu_orca_2022}, paged attention~\cite{kwon_efficient_2023}, prompt processing disaggregation~\cite{strati_dejavu_2024,patel_splitwise_2024,zhong_distserve_2024}, and tensor parallelism~\cite{li_alpaserve_2023} optimize inference latency and throughput for individual models. However, the growing need to serve many model variants concurrently presents additional challenges. The need is particularly pronounced for LLM service providers, such as DeepInfra~\cite{deepinfracustommodels}, Together.AI~\cite{togetheraicustommodels}, or Fireworks.AI~\cite{fireworksondemand}, who serve not only the popular base model, but also various fine-tuned models from either customers or the community.

On these service providers, the popularity of each model varies significantly between base models and their fine-tuned variants.
From OpenRouter statistics~\cite{openrouter_stats}, while the base model (\textit{Llama-3.1-70B}) consumes $\sim$1 billion tokens per day, the deployed fine-tuned variants consume much less, e.g., $\sim$100 million for \textit{Nemotron-70B} and \textit{Hermes-3-70B}, and less than 10 million for \textit{Sonar-70B} and \textit{Lumimaid-70B}. In addition, Figure~\ref{fig:motivation_trace} shows that requests to base models and their variants can range from sporadic to very dense. For example, the models \textit{alpaca-13b}~\cite{alpaca_model}, \textit{koala-13b}~\cite{koala_model}, \textit{gpt4all-13b}~\cite{gpt4all_model}, \textit{vicuna-13b}~\cite{vicuna_model} and \textit{wizardlm-13b}~\cite{wizardlm_model} are all fine-tuned variants of the base model \textit{llama-13b}. We observe how certain variants, such as \textit{wizardlm-13b} and \textit{gpt4all-13b}, are highly popular over time while others, such as \textit{alpaca-13b} and \textit{vicuna-13b}, have sporadic requests. The varying popularity may be due to different accuracy on certain downstream tasks, and the irregular request patterns make it difficult to batch requests and decide how to provision GPU resources for each model.

Dedicating GPUs for each model variant minimizes latency, but requires many expensive GPUs that would sit idle for long periods of time (yellow areas in Figure~\ref{fig:motivation_trace}).
On the other hand, swapping model variants on a limited pool of GPUs reduces cost and improves utilization, but adds latency on the critical path of requests.
Optimizing the storage hierarchy to reduce swapping latency can help~\cite{fu2024serverlessllm}, but requests still experience long queuing delays due to limited batching opportunities when treating each model variant as a separate model.

For LLM service providers, serving this long tail of less popular models cost-efficiently while maintaining low latency and high quality is a key challenge. The state-of-the-art approach to LLM serving with the proliferation of fine-tuned model variants is to adopt an entirely new fine-tuning paradigm, \textit{parameter-efficient tuning} (PEFT). Instead of tuning model weights directly, PEFT tunes a compact, adjunctive model adapter (i.e., a small set of extra parameters) that is attached to the model for serving. For example, low-rank adaptation (LoRA)~\cite{hu_lora_2021} is a popular PEFT method that freezes models weights and attaches low-rank matrices to the model structure which are fine-tuned on task-specific data. Systems like Punica~\cite{chen2024punica} and S-LoRA~\cite{sheng_s-lora_2023} leverage the small size of LoRA adapters and the common base model weights to efficiently swap adapters and batch requests to optimize inference latency and throughput. Given the needs for deploying fine-tuned LLMs, these solutions have been integrated into commercial platforms such as LoRAX~\cite{predibase_lorax}.

\begin{figure}[t]
  \centering
  \includegraphics[width=\linewidth]{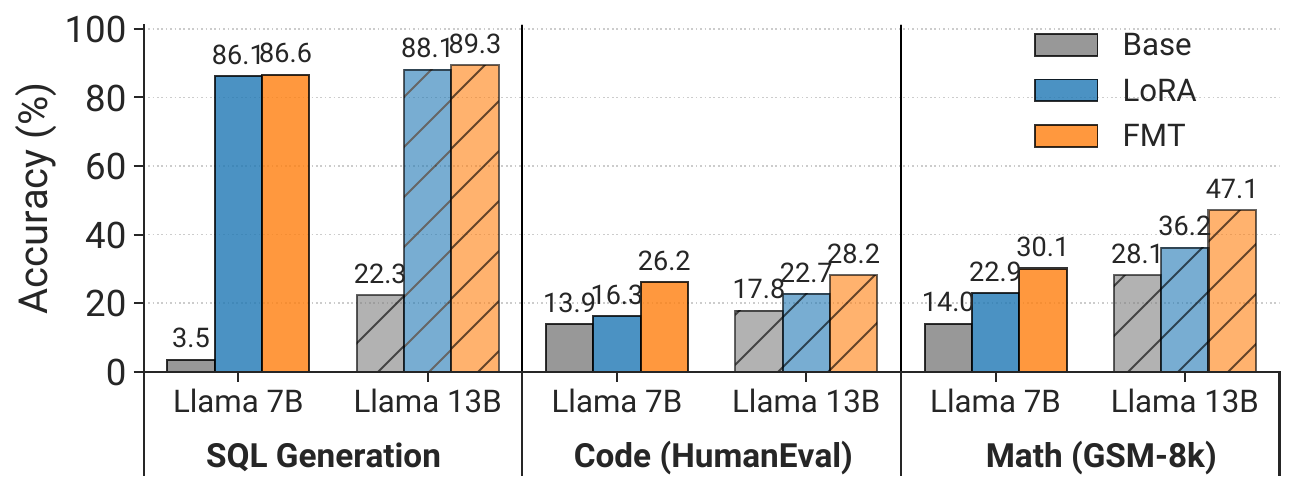}
  \caption{LoRA vs. full-model fine-tuning accuracy~\cite{anyscale_fine-tuning_2023,lora-learns-forgets-biderman2024lora}. LoRA fine-tuning is comparable for some tasks (SQL), but has lower quality on more complex tasks (Math and Code).}
  \label{fig:motivation_compare_lora}
\end{figure}

However, PEFT serving systems are not compatible with the traditional fine-tuning approach, \textit{full model tuning} (FMT). While PEFT methods have achieved high accuracy for downstream tasks like SQL generation~\cite{b-mc2_2023_sql-create-context,anyscale_fine-tuning_2023} and ViGGO~\cite{juraska_viggo_2019}, they are still not able to match the accuracy of FMT for more complex tasks, such as coding and math~\cite{lora-learns-forgets-biderman2024lora}, or when the fine-tuning dataset is particularly large~\cite{zhang_when_2024}. Figure~\ref{fig:motivation_compare_lora} summarizes the results of two recent studies~\cite{anyscale_fine-tuning_2023,lora-learns-forgets-biderman2024lora} comparing LoRA and FMT accuracy on three downstream tasks. Full-parameter fine-tuning remains appealing to applications that aim to maximize accuracy on more complex tasks and is still widely-used. Yet the serving solutions available for this paradigm (which either dedicate GPUs for each model variant or swap entire models) are either expensive or slow. 

We aim to extend PEFT-based serving systems to also support efficient FMT model variant serving.
Our key insight is that FMT model weights often have low-magnitude perturbations with respect to the original pre-trained model (see Figure~\ref{fig:weight_distribution}), allowing us to aggressively sparsify, quantize, and compress \textit{model deltas} while maintaining high accuracy. 
Due to their compact size,  low-precision and sparse deltas can be swapped and served with low latency. We apply this idea in the design of \projectname, a multi-variant LLM serving system. \projectname extracts and compresses model deltas with \algname, an algorithm we propose to help maintain high accuracy during compression. To efficiently serve FMT model variants, \projectname decouples base model serving and low-precision delta serving, 
inspired by how S-LoRA~\cite{sheng_s-lora_2023} and Punica~\cite{chen2024punica} serve LoRA adapters. This decoupling enables \projectname to batch requests to different model variants that share the same base and perform low-precision inference for model deltas to minimize latency and memory bandwidth pressure on the GPU.
We further optimize delta inference by designing a custom GPU kernel, \underline{S}elective \underline{B}atched \underline{M}atrix \underline{M}ultiplication (\textit{SBMM}), which selectively batches requests to the same delta to minimize random accesses and performs operations for multiple deltas in parallel to amortize kernel launch overhead.
We build \projectname on top of vLLM~\cite{kwon_efficient_2023} (which supports S-LoRA/Punica-based LoRA-serving~\cite{chen_punica_2023,sheng_s-lora_2023}) and adapt continuous batching~\cite{yu_orca_2022} and model parallelism~\cite{shoeybi_megatron-lm_2020} for delta-based model serving. 

To the best of our knowledge, \projectname is the first serving system to support both FMT and PEFT model variants while accelerating FMT model serving with hardware-optimized delta compression.
In summary, our key contributions are:
\begin{itemize}[leftmargin=*, itemsep=0pt, topsep=0pt]
  \item We propose \algname (\S\ref{sec:delta_compression_pipeline}), a \textit{hardware-efficient} compression algorithm that aggressively compresses model deltas post full-model fine-tuning. It applies structured sparsity, quantization, and optionally lossless compression. \algname can compress a 70B-parameter Llama-2 model delta by 13$\times$ while maintaining comparable accuracy to the original FMT model. In contrast, applying a similar sparsification and quantization technique like SparseGPT~\cite{frantar_sparsegpt_2023} directly on the fine-tuned model substantially degrades accuracy, even with only 6$\times$ compression.
  \item We design \projectname (\S\ref{sec:system_design}), a serving system that leverages \algname to efficiently serve FMT model variants. It decouples and parallelizes base and delta model inference, reducing queuing delays by batching requests to different model variants derived from the same base and leveraging our custom GPU kernel for hardware-optimized low-precision and sparse delta serving. 
  \projectname achieves 2$\times$ to 12$\times$ higher throughput compared to vLLM~\cite{kwon_efficient_2023}.
  \item We identify challenges when deploying \projectname in practice and propose solutions to address them. In particular, we study how many deltas to place concurrently to balance GPU memory usage and how to reduce starvation when serving many model deltas.
\end{itemize}

Our compression pipeline and serving system is open source and available at \url{https://github.com/eth-easl/deltazip}.

\section{Background and Motivation}
\label{sec:background}

We first introduce key concepts in LLM fine-tuning and serving, then highlight key challenges and opportunities. 

\subsection{Background}
\label{sec:background_language_modeling}

\textbf{LLM fine-tuning.} 
In contrast to prior deep learning workloads that follow a \emph{task-specific} paradigm, which trains a model from scratch on domain-specific data to tackle a particular task (e.g., machine translation~\cite{wu2016google, sutskever2014sequence}), LLMs follow a \emph{pretrain-then-finetune} paradigm, which pretrains a model on a massive general-domain corpus and then fine-tunes it for specific objectives. For example, ChatGPT is fine-tuned to follow human instructions~\cite{ouyang_training_2022}. There are two main fine-tuning approaches: 
1) {\emph{Full Model Tuning} (FMT)} which updates all model parameters and 
2) {\emph{Parameter-Efficient Tuning} (PEFT)} which adds a small number of extra parameters after pretraining, called \textit{adapters}, e.g., low-rank matrices learned during fine-tuning. PEFT methods, such as LoRA~\cite{hu_lora_2021}, are popular ways to reduce the compute and memory requirements of both fine-tuning and serving. However, the choice of fine-tuning paradigm impacts accuracy. While PEFT methods can achieve high accuracy for a variety of tasks~\cite{hu_lora_2021}, recent studies~\cite{lora-learns-forgets-biderman2024lora, zhang_when_2024,anyscale_fine-tuning_2023} --- summarized in Figure~\ref{fig:motivation_compare_lora} --- reveal that FMT still achieves higher accuracy for more complex tasks. 

\textbf{LLM compression.}
Model compression is a popular approach to reduce the memory and compute requirements of LLM inference in resource-constrained environments. Techniques like GPTQ~\cite{frantar_gptq_2023}, SparseGPT~\cite{frantar_sparsegpt_2023} and AWQ~\cite{AWQ} reduce memory footprints and improve latency while maintaining model quality (when applied in moderation). Pushing these techniques to extremely low bit-width quantization and sparsity, such as 2-bit quantization and more than 50\% sparsity, results in significant model quality degradation~\cite{chai_int21_2023, liu_emergent_2023}.

\textbf{LLM serving.} 
The LLM inference involves two phases: 
(1) {\emph{prompt processing} (prefill)}, where the tokens (i.e., basic units of text) in the input prompt are processed in parallel. This phase can be parallelized since all previous tokens are known from the user-provided prompt, and it is usually compute-bound.
(2) {\emph{token generation} (decode)}, where the model iteratively generates one token for each forward pass. Due to the inter-token data dependency, this phase cannot be parallelized and is typically memory-bound. Token generation stops when the model generates an end-of-sequence (EOS) token or meets a user-defined condition (e.g., reach the maximum number of generated tokens). Many works~\cite{yu_orca_2022,kwon_efficient_2023,chen2024punica}, including our own, focus on optimizing the token generation phase as it is the main bottleneck for LLM serving.

\subsection{Challenges for Multi-Variant Model Serving}
\label{sec:bg_problems}

With the proliferation of fine-tuned LLM variants, each specialized for a particular user's task or domain, it is critical to design LLM serving systems that can efficiently multiplex requests to different model variants. Yet, on today's commercial LLM serving platforms, fine-tuned model variant serving is still more expensive than base model serving~\cite{llmcost, openai_openai_nodate}. The high cost is due to several key challenges that lead to low resource utilization:

\textbf{Low request rates per model variant.} A naive approach to serve multiple fine-tuned models is to consider each model as a separate model and dedicate a group of GPUs for each variant. However, as shown in Figure~\ref{fig:motivation_trace}, the requests to each model variant are sporadic and often low in volume, which limits the batch size. Hence, dedicating GPUs for each model underutilizes resources and leads to high cost.

\textbf{Swapping incurs high latency.} Another approach is to swap the models in and out of a limited pool of GPUs. However, the unpredictable nature of model invocations makes it hard to predict when a request for a particular model will arrive to enable prefetching. Swapping models in and out of GPUs on the critical path of requests leads to high latency and GPUs also remain underutilized.

\textbf{Accuracy gap between PEFT and FMT.}
Although PEFT methods such as LoRA~\cite{hu_lora_2021} produce significantly smaller fine-tuned adapters and systems~\cite{sheng_s-lora_2023,chen_punica_2023,wu2024dlora,zhou_pets_2022} have been built for serving adapters, these methods face the challenge of model quality. As shown by recent studies~\cite{lora-learns-forgets-biderman2024lora, zhang_when_2024,anyscale_fine-tuning_2023} (summarized in Figure~\ref{fig:motivation_compare_lora}), FMT still achieves higher accuracy for more complex tasks compared to PEFT methods.
\section{\projectname Overview}

\begin{figure}[t]
      \centering
      \includegraphics[width=\linewidth]{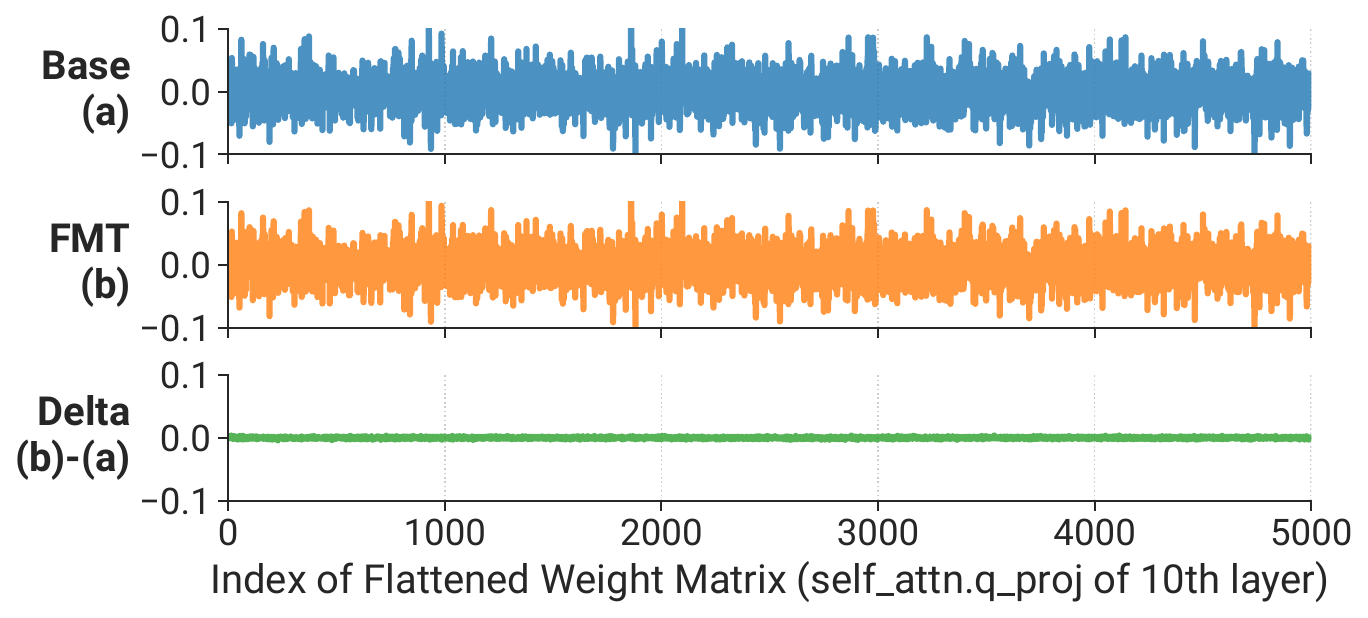}
      \caption{Flattened weight matrix in an intermediate layer of the pre-trained model (a), the fine-tuned model (b), and the model delta between the two (bottom, (b)-(a)).}
      \label{fig:weight_distribution}
\end{figure}

We observe three key opportunities to address the aforementioned challenges and improve the efficiency of FMT serving (\S\ref{sec:opportunities}). We then describe how we leverage these opportunities to design \projectname, a multi-variant LLM serving system that complements parameter-efficient model serving with hardware-efficient full model tuning serving (\S\ref{sec:system_arch}).

\subsection{Opportunities and Key Insights}
\label{sec:opportunities}

\textbf{1. Model deltas are highly compressible}.
Although fine-tuning can significantly improve model performance on specific tasks, we find it typically results in small-magnitude changes to model parameters. Figure~\ref{fig:weight_distribution} shows the distribution of the weight matrix $\mathbf{w_q}$ in a transformer layer of a pre-trained \texttt{Llama-2-7b} model, its fine-tuned counterpart \texttt{Vicuna-7b-v1.5} model, and the model delta, which is obtained by subtracting the base model from the fine-tuned model. The delta has significantly smaller magnitude range and fewer outliers. This makes the delta easier to compress than the model itself, with both quantization~\cite{AWQ, frantar_gptq_2023} and sparsification~\cite{frantar_sparsegpt_2023}. As quantization involves calculating the maximum value such that the quantization preserves the outliers in the weight matrix, a more concentrated distribution results in a denser quantization grid, which can maintain model quality while using lower bit width. In addition, the prevalence of near-zero values makes it easier to apply sparsification to the model delta than to the full model.
We will show that this enables us to compress model deltas by over 10$\times$, making it fast to swap and serve while achieving comparable quality to full-precision model serving.

\textbf{2. Many model variants share the same base model, for which we can batch many requests}.
Since pre-training LLMs requires immense datasets and compute resources,  only a handful of organizations are well-positioned to produce high-quality pre-trained models~\cite{llama3,gemini_2024}. Instead, most organizations rely on fine-tuning pre-trained models for their use cases, as this is typically much cheaper~\cite{llmcost}. This means that fine-tuned models often share the same base model, even if they are used for different applications. For example, GitHub's Copilot~\cite{copilotx} and OpenAI's ChatGPT~\cite{ouyang_training_2022} are both fine-tuned from GPT models. Hence, although batch sizes for individual model variants may be limited (Figure~\ref{fig:motivation_trace}), we can quickly accumulate large batches to common base models to improve GPU utilization and reduce queuing delays.  This requires decoupling base model and delta serving, which we will discuss in \S\ref{sec:base-and-delta-decoupling}.

\textbf{3. GPU support for delta computations}.
We can accelerate delta inference by leveraging features of modern GPU hardware, such as:
1) using sparse tensor cores~\cite{bai_structured_2023} for sparse delta computations, 
2) performing multiple low-precision matrix multiplications in parallel to improve stream multi-processors (SMs) utilization, and
3) reducing memory bandwidth pressure for delta matrix computations by designing a custom GPU kernel that minimizes data movement from GPU global memory to device memory.

\subsection{System Overview}\label{sec:system_arch}

To realize the above opportunities in practice, we propose \projectname. 
Our key contributions are a model delta compression algorithm (\textit{\rawalgname}) and the end-to-end \projectname serving system. The algorithm compresses FMT deltas into a hardware-efficient, low-precision, and sparse format that preserves high accuracy, while the serving system incorporates state-of-the-art serving optimizations and adapts them for low-latency, high-throughput inference with model deltas. \projectname is designed for LLM service providers to operate, and LLM users can interact with the system through the standard HTTP API frontend, as if using a dedicated LLM.

\begin{figure}[tp]
  \centering
  \includegraphics[width=\linewidth]{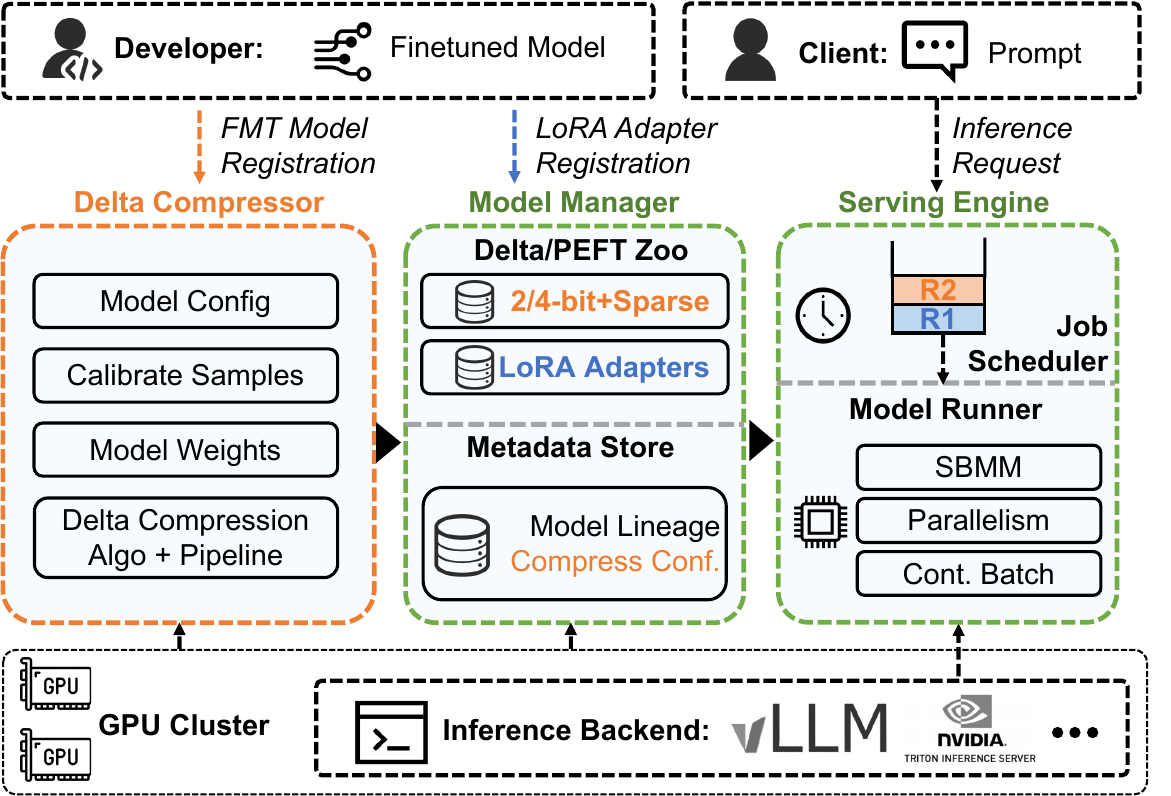}
  \caption{\projectname system architecture.}
  \label{fig:sys_arch}
\end{figure}

\textbf{System architecture.} Figure~\ref{fig:sys_arch} shows the  \projectname system architecture. 
The system consists of three main components: 1) the \textit{Delta Compressor} (\S\ref{sec:delta_compression_pipeline}) which extracts and compresses the delta from a FMT model registered by the user, 2) the \textit{Model Manager} which tracks and manages model metadata, and 3) the \textit{Serving Engine} (\S\ref{sec:system_design}) which serves inference requests for base and fine-tuned models.

\begin{figure*}[th]
\vspace{-10pt}  
    \includegraphics[width=\textwidth]{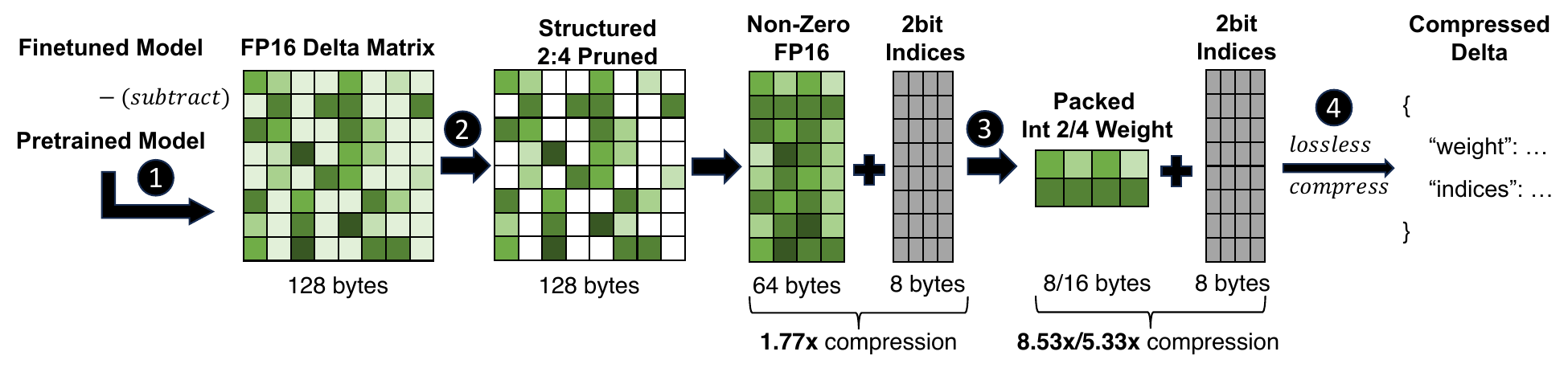}
    \caption{\projectname Compression Pipeline. The pipeline consists of delta extraction, sparsification \& quantization, and optionally lossless compression. The compressed delta is stored as a dictionary of compressed weight matrices and metadata.}
    \label{fig:compression_pipeline}
\end{figure*}

\textbf{Life of a request.}
The model developer first uploads a fine-tuned model to the Delta Compressor, together with some metadata (such as the pre-trained model identifier) and a small calibration dataset that the compression algorithm uses to measure and minimize accuracy loss. The compressor computes and compresses the model delta, then stores it in a packed, low-precision, and sparse format in the Model Manager's delta zoo. The manager keeps track of metadata for each stored delta, such as its compression configuration (such as the bit width per parameter and sparsity level) and model lineage. In addition to deltas, the model manager also allows developers to register LoRA adapters directly, and the serving engine can serve them as well.

The Serving Engine serves inference requests for fine-tuned models whose deltas are stored in the model manager. Users send inference requests to the engine's API frontend, which fetches the requested deltas into CPU main memory, if they are not already present in CPU or GPU memory. Meanwhile, the frontend forwards the request to the job scheduler, which queues the requests. The job scheduler performs continuous batching by assigning requests to the model runner per-iteration (i.e., for each forward pass of the model). The model runner is responsible for processing the batched requests. Internally, the model runner can process requests for different FMT and PEFT models in parallel, by decoupling base model and delta (or PEFT adapter) inference computations. The model runner also leverages tensor parallelism and supports large models that do not fit in a single GPU. 

\begin{figure}[tp]
  \includegraphics[width=\linewidth]{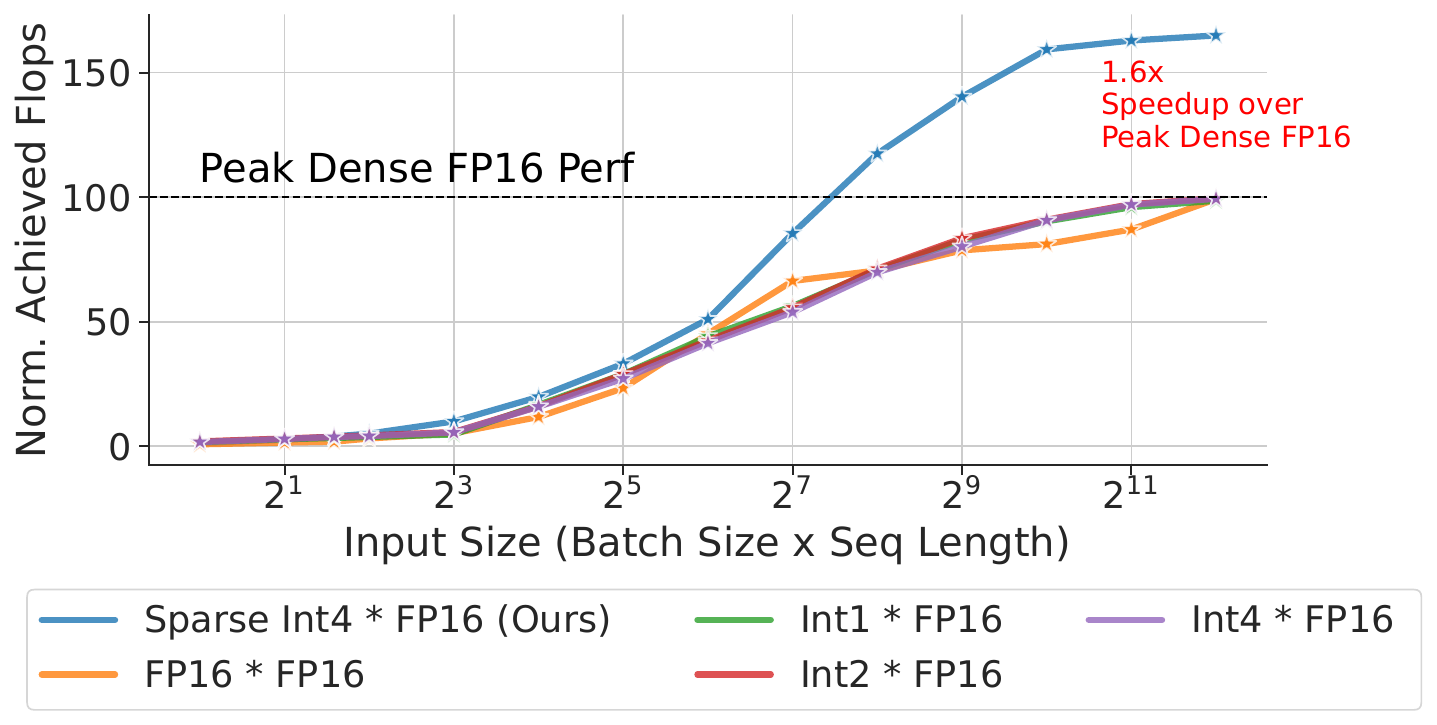}
  \caption{(Compressed) Matrix Multiplication Performance}
  \label{fig:matmul_microbench}
\end{figure}

\section{Model Delta Compression}
\label{sec:delta_compression_pipeline}

When model developers register an FMT model into the system, the Delta Compressor computes the delta and applies a compression pipeline (\S\ref{compression-pipeline}) to reduce the size of the model state. The compression pipeline only runs when a developer registers a new model. We develop the \algname algorithm (\S\ref{compression-alg}) to apply quantization and sparsification in the compression pipeline in a way that preserves high accuracy. 

We design \projectname to be agnostic to the compression pipeline, such that users can apply a variety of compression techniques to model deltas, such as GPTQ~\cite{frantar_gptq_2023}, SparseGPT~\cite{frantar_sparsegpt_2023}, and AWQ~\cite{AWQ}. Here we describe a SparseGPT-inspired compression pipeline, which we use in our implementation and evaluation of \projectname. We implemented this compression pipeline on top of AutoGPTQ~\cite{panqi_william_wei_panqiweiautogptq_2023} and SparseGPT~\cite{frantar_sparsegpt_2023}.

\subsection{\projectname Compression Pipeline}\label{compression-pipeline}

Within \projectname, the Delta Compressor encompasses multiple steps as shown in Figure~\ref{fig:compression_pipeline}. Initially, Step 1 computes the model delta by subtracting the pre-trained base model weights from the fine-tuned model. The remaining steps are applied on the delta matrix since it has a smaller range of values (see Figure~\ref{fig:weight_distribution}) and is hence more amenable to compression compared to the fine-tuned model weight matrix.

Step 2 applies structured 2:4 pruning~\cite{zhou_learning_2021,hubara_accelerated_2021}, which involves setting at least two elements among each group of four contiguous elements to 0 in the delta matrix. With structured pruning, \projectname only stores the non-zero values of the delta matrix and their 2-bit indices. The main motivation for applying structured sparsity is due to its \textit{hardware-efficiency}. Compared to quantization only or unstructured sparsity, the particular pattern we used in structured sparsity enables higher peak performance with large input size, since modern NVIDIA GPUs from the Ampere architecture onwards and AMD RDNA/CDNA GPUs have specialized hardware support for this feature~\cite{bai_structured_2023, amdsparsitysupport}. This hardware support is automatically enabled when running compatible kernels. Given the trend toward hardware optimization, we expect future GPUs will continue to support this feature.

Figure~\ref{fig:matmul_microbench} shows a microbenchmark of matrix multiplication performance with different compression configurations. We observe that with small input sizes (e.g., from 1 to 4, which is common during the decode phase), structured sparse matrix multiplication achieves similar performance to quantization-only compression and outperforms the uncompressed version. This is mainly because both sparsity and quantization reduce data movement between GPU HBM and compute unit. However, with large input sizes (e.g., from 16 to 4096, which is typical during the prefill phase), structured sparsity significantly outperforms quantization-only compression, as structured sparsity leverages the sparse tensor cores~\cite{bai_structured_2023} on GPUs to achieve higher performance.

Step 3 quantizes the pruned delta weight matrix, which squeezes the values into a smaller bit-width format and packs the data. For example, 4-bit quantization packs 8 values into a single 32-bit value, achieving 4$\times$ compression ratio.

Step 4 is an optional final step that applies lossless compression. We use the GDeflate algorithm from \texttt{nvcomp}~\cite{nvidia_nvcomp_2021} for lossless compression for fast decompression on GPUs. This step is beneficial when the disk bandwidth is a bottleneck (such as with NFS). In such cases, users can opt-in to lossless compression to reduce the disk I/O time. If disk I/O is not a bottleneck, lossless compression may not be beneficial due to the decompression overhead.

\textbf{Hardware-efficient design.} We design \projectname's compression pipeline with GPU hardware features in mind. Structured pruning (Step 2) leverages sparse tensor cores~\cite{bai_structured_2023} for fast sparse matrix multiplication. The quantization (Step 3) allows us to move a smaller amount of data between GPU global memory to device memory, alleviating the memory bandwidth bottleneck. Lossless compression (Step 4) uses GPU decompression engines~\cite{nvidia_nvcomp_2021}.

\subsection{\algname Algorithm}\label{compression-alg}

The core of the compression pipeline is the lossy compression algorithm that finds the optimal pruning mask (in Step 2) and quantized weight matrix (in Step 3).
We design \algname to compress the model deltas in a way that minimizes the loss between the outputs computed by the original weights and the compressed weights. \algname achieves this by calibrating the compression algorithm with a subset of the training dataset, provided by model developers when registering the FMT model to \projectname. As outlined in Algorithm~\ref{alg:fmzip_obs}, \algname iteratively (Line~\ref{alg_line1}) compresses the delta (extracted in Line~\ref{alg_line4}) for each layer of the model. For each layer, the objective is to find the optimal pruning mask $\mathbf{M}$ and quantized weight matrix $\mathbf{Q}$ (Line~\ref{alg_line2}, Line~\ref{alg_line3}). \algname is designed to be compatible with various different compression techniques to achieve this goal, such as GPTQ~\cite{frantar_gptq_2023}, SparseGPT~\cite{frantar_sparsegpt_2023} and AWQ~\cite{AWQ}. In our current implementation, we follow the optimal brain surgeon~\cite{lecun_optimal_1989,hassibi_optimal_1993} approach and leverage SparseGPT~\cite{frantar_sparsegpt_2023} (since it has support for both sparsification and quantization) to compute the optimal $\mathbf{M}$ and $\mathbf{Q}$ by solving the following optimization problem:

\begin{algorithm}[t]
    \caption{\algname algorithm. Given an $N$-layer FMT model where each layer has a weight matrix $\mathbf{w}_f$ of the shape $d_\text{row}\times d_\text{col}$, and the corresponding layer in the base model has a weight $\mathbf{w}_b$, the algorithm computes quantized weight $\mathbf{Q}$ and pruning mask $\mathbf{M}$. $\odot$ denotes elementwise multiplication.}\label{alg:fmzip_obs}
    \begin{algorithmic}[1]
    \For{$n=0,1,2,\cdots, N$}
    \label{alg_line1}
    \State $\mathbf{M} \gets \mathbf{1}_{d_{\text{row}}\times d_{\text{col}}}$ \Comment{Binary Pruning Mask}
    \label{alg_line2}
    \State $\mathbf{Q} \gets \mathbf{0}_{d_{\text{row}}\times d_{\text{col}}}$ \Comment{Quantized Delta}
    \label{alg_line3}
    \State $\Delta = \mathbf{w}_f \textcolor{teal}{- \mathbf{w}_b}$\Comment{Extract Delta}
    \label{alg_line4}
    \State $\mathbf{Q}, \mathbf{M} = \text{Compress}(\Delta, \mathbf{X}_n)$ \Comment{e.g., SparseGPT}
    \label{alg_line5}
    \State $\tilde{\mathbf{w}}_q \gets \mathbf{Q} \odot \mathbf{M} \textcolor{teal}{ + \mathbf{w}_b}$ \Comment{Reconstruct Weight}
    \label{alg_line6}
    \State $\mathbf{X}_{n+1}=\tilde{\mathbf{w}}_f \mathbf{X}_n$ as input of next layer.
    \label{alg_line7}
    \EndFor
    \For{$n=0,1,2\cdots, N$}
    \State pack and store $\mathbf{Q},\mathbf{M}$ of layer $n$.
    \EndFor
    \end{algorithmic}
\end{algorithm}

\begin{equation}
\argmin_{\tilde{\Delta}} ||\Delta\cdot\mathbf{X} - \tilde{\Delta}\cdot\mathbf{X}||_2^2
\end{equation}
where $\tilde{\Delta}$ is the compressed delta and $\mathbf{X}$ is the input to the layer, which is from the calibration set used for compression.

The major distinction from full model compression is that \algname reconstructs the weight matrix for each layer (Line~\ref{alg_line6}) after compressing the delta and computes the input for next layer (Line~\ref{alg_line7}). 
This is because the input data (i.e., $\mathbf{X_n}$ in Line~\ref{alg_line5}) is crucial for the compression algorithm. 
Without re-adding the base weight matrix and reconstructing the weight matrix for each layer, the magnitude of the deltas causes diminishing outputs, leading to vanishing activations in deeper layers. The vanishing activations, as the input to the next layer, will make the compression algorithm fail to capture the input (i.e., $X_n$ Line~\ref{alg_line5}). By extracting the delta and reconstructing the weight for each layer on the fly, \algname ensures proper calibration and maintains high model quality.

Beyond model quality, \algname has three advantages: 1) \textbf{Low memory requirement}. As \algname performs layer-wise compression, it only needs a GPU with sufficient memory for a single layer to perform the forward pass. \newline 2) \textbf{No need to retrain the model}. Unlike some other compression algorithms~\cite{chai_int21_2023,tseng2024quip}, \algname does not require further fine-tuning the model to recover the model quality. 3) \textbf{Data-Efficiency.} \algname requires a small calibration set to achieve high accuracy, which does not need to be the exact, entire training set. In our experiments, we found $256$ samples chosen from open-source instruction tuning dataset \texttt{UltraChat}~\cite{ding2023enhancingchatlanguagemodels} are sufficient for calibration. In practice, if the users fine-tune their models through the service provider, then the training data can be directly used as the calibration set. If the users fine-tune their models on their own, they can use a small, representative subset (e.g., by uniform sampling) of the training data as the calibration set. The users can also evaluate the model quality after compression using an evaluation set to ensure the quality meets their requirements, and adjust the calibration set if needed.

Depending on the model size and hardware, the compression process takes some time to run. From our experiments, compressing a 7B model takes around 30 minutes on a single RTX 3090 GPU and AMD EPYC 7313 16-Core CPU. Although this still introduces extra cost and overhead, the process is offline (i.e., not on the critical inference path) and only runs once when the model is registered, making it acceptable.
\section{Serving System Design}
\label{sec:system_design}
\baselineskip=12.1pt

We implement \projectname in 18K lines of Python and 1K lines of C++/CUDA code. We build the serving engine on top of vLLM~\cite{kwon_efficient_2023}, HuggingFace Transformers~\cite{wolf_transformers_2020},  and Sparse Marlin~\cite{castro2024sparsemarlin}.
\S\ref{sec:base-and-delta-decoupling} explains how \projectname decouples base and delta serving to maximize request batching for base model inference. \S\ref{sec:gpu-delta-serving} describes how the system parallelizes low-precision delta serving with a custom GPU kernel design. Finally, we describe how we extend model parallelism for delta serving (\S\ref{sec:model-parallelism}) and schedule delta serving requests with continuous batching (\S\ref{sec:continuous-batching}).

\subsection{Base and Delta Decoupling}
\label{sec:base-and-delta-decoupling}
 
\projectname's serving engine always keeps the base model in GPU memory and swaps compressed deltas on-demand to serve inference requests. The naive approach to serve a fine-tuned model is to load its compressed delta, decompress it, add it to the base model, and then perform inference. However, this approach is inefficient for several reasons: 1) it requires decompressing the delta on the critical path, which adds latency; 2) it does not allow batching requests to different model variants with the same base model; 3) performing inference after adding the delta back to the full-precision base model does not leverage low-precision computation to reduce delta inference latency; and 4) storing decompressed deltas in GPU memory to add back to the base model limits how many deltas can fit concurrently in the GPU.

To improve inference latency and throughput, our first optimization is to \textbf{decouple the inference computation of the base model and delta}. Consider a matrix multiplication, which we can decouple as follows by the distributive law:
\begin{equation}\label{eq:distributive_law}
  \begin{aligned}
      \mathbf{Y} & =\mathbf{w}_\text{fine-tuned}\mathbf{X} =(\mathbf{w}_\text{base}+\Delta)\mathbf{X}\\
                 & \approx\underbrace{\mathbf{w}_\text{base}\cdot\mathbf{X}}_{\text{Batched FP16 matmul}} + \underbrace{\Delta\cdot\mathbf{X}}_{\text{Quantized and sparse matmul}}
  \end{aligned}
\end{equation}

In Eq~\ref{eq:distributive_law}, $\mathbf{w}_\text{base}\mathbf{X}$ refers to the matmul with the base model, which is shared across all fine-tuned models and we can compute this with a standard \texttt{GEMM}. $\Delta\mathbf{X}$ denotes the computation with the delta, which is in a sparse and low-precision format. \projectname decouples the computation into two parts and executes the batched base model and low-precision delta matrix multiplications independently and in parallel.

Since the distributive law does not hold for non-linear operations, such as activation functions, we decouple the computation at the granularity of linear layers. We merge results from the base model and the delta part after each linear layer to get the output to feed into a non-linear operation. In a transformer block, we serve all linear layers with low-precision, such as the QKV projections $\mathbf{w_q}, \mathbf{w_k}, \mathbf{w_v}$, output projection $\mathbf{w_o}$ and the linear layers in the MLP module. 

Decoupling base and delta serving improves GPU utilization and performance in several ways. First, for the base model computation, it enables batching requests for different model variants, as long as they share the same base model. Second, for the delta computation, keeping deltas in low-precision, sparse formats allows us to fit more deltas in GPU memory and reduce swapping. Third, low-precision delta serving also reduces inference latency since token generation in LLM inference is inherently memory-bound (as discussed in \S\ref{sec:background_language_modeling}), so the decoding latency is proportional to the GPU memory consumption of the model weights.

If there are $M$ base models and $M>1$, we divide the GPU cluster into $M$ sets of GPUs, each dedicated to serving a particular base model and its fine-tuned variants. Hence, even with many model variants, we only need $M$ sets of GPUs to serve all these models. LoRA serving systems~\cite{chen_punica_2023,sheng_s-lora_2023} have a similar assumption. However, operators can also deploy multiple base models per GPU if there is sufficient memory.

\subsection{GPU-Efficient Delta Serving}\label{sec:gpu-delta-serving}

\begin{figure}[tp]
  \includegraphics[width=\linewidth]{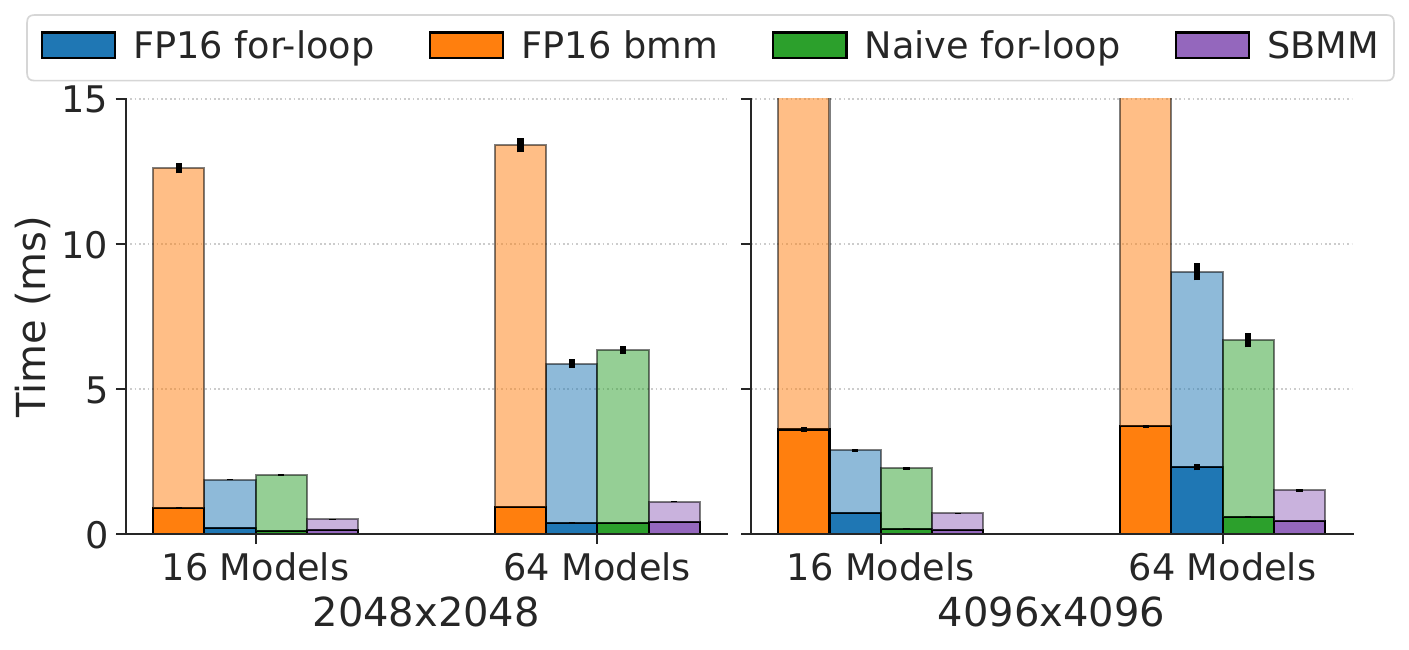}
  \caption{Breakdown of total execution time for different implementations of batched matrix multiplication. The dark part of the bar shows the portion of total execution time spent on computation. Naive for-loop refers to performing low-precision and sparse computation while looping through all models. SBMM refers to our proposed kernel.}
  \label{fig:kernel_launch_overhead}
\end{figure}

\projectname parallelizes the delta computation ($\Delta X$) in a GPU-efficient manner with an approach we call \underline{S}elective \underline{B}atched \underline{M}atrix \underline{M}ultiplication (\textit{SBMM}).
For a single batch of requests  $\mathbf{X}_i^j$ where $i$ is the request index and $j$ is the delta index, the delta computation can be formalized as: given a batch of requests $\mathbf{X}_1^{1}, \mathbf{X}_2^{2}, \ldots, \mathbf{X}_i^{j}$, compute $\mathbf{Y}_1=\mathbf{X}_1^{1}\cdot \mathbf{\Delta}_{1},\ldots, \mathbf{Y}_i=\mathbf{X}_i^{j}\cdot\mathbf{\Delta}_{j}$. We use an index $Idx_i$ to denote the delta index for each request $i$.
Naively, we can loop through different deltas in the batch, find the respective requests and compute the matrix multiplication. However, we observe that this approach is inefficient for two main reasons: 1) it introduces random memory accesses to fetch the inputs and write the outputs to the correct locations, 2) it needs to launch the matrix multiplication kernel multiple times which computes on a small number of requests each time, incurring a high overhead and low GPU utilization. Another option is to use an operator with a batch dimension like \texttt{torch.bmm}~\cite{pytorch_foundation_torchbmm_nodate}. However, this requires first stacking the weight matrices for each input into a single matrix, which is not efficient or scalable with delta matrices, as they have large memory footprints. Figure~\ref{fig:kernel_launch_overhead} shows the total execution time of different batched matrix multiplication implementations and the portion of time each spends on computation. We find that although the low-precision matmul kernel reduces computation time, the total execution time is still high as it is dominated by kernel launch time and other overheads.

\begin{figure}[tp]
  \includegraphics[width=\linewidth]{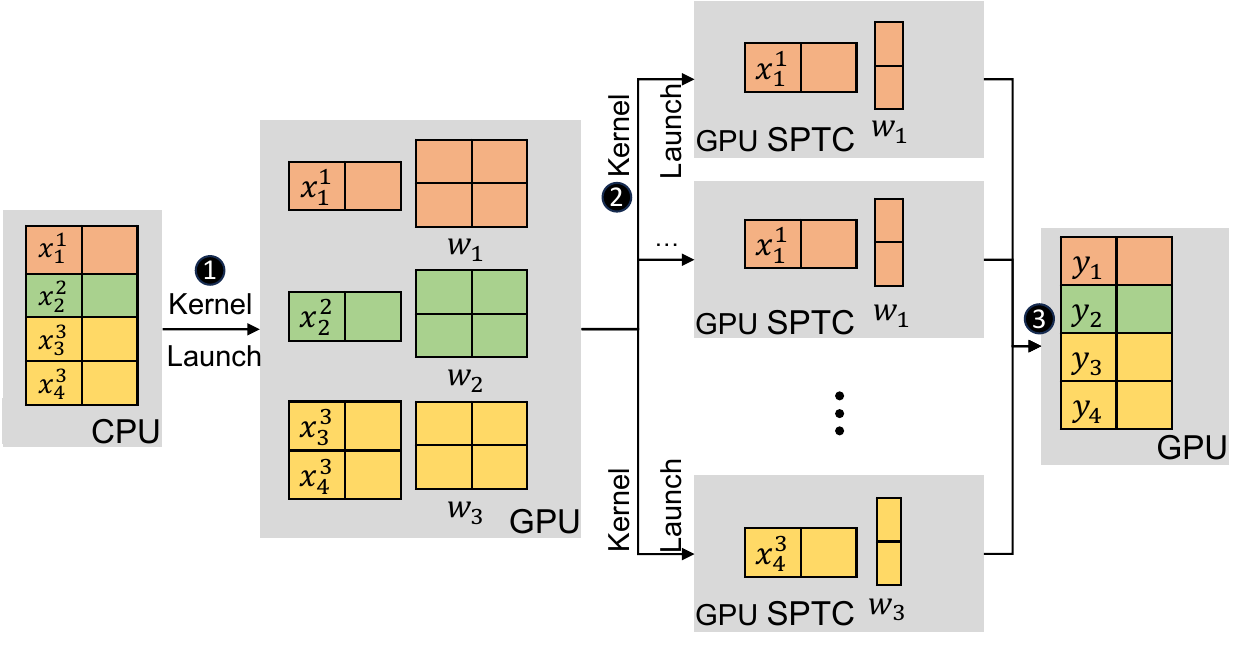}
  \caption{SBMM kernel launch. The kernel computes the output $y_i = \text{matmul}(x_i, w_{Idx_i})$ where the matmul is low precision and sparse. Different colors represent different deltas and their respective requests. SPTC=sparse tensor cores. }
  \label{fig:kernel_launch}
\end{figure}

To reduce this overhead and improve GPU utilization, we apply two optimizations. First, before launching a computation on the GPU, \projectname's job scheduler \textbf{reorders the requests to group requests belonging to the same delta together}. This reduces random data accesses during computation and enables higher batch sizes for delta serving.

Second, we \textbf{design a kernel that performs the SBMM operation for multiple deltas in a single kernel launch}.
We design SBMM to compute its kernel launch configuration dynamically to balance resources for each delta, since there is often a different number of requests for each delta in the batch. We implement this with CUDA dynamic parallelism~\cite{andy_cuda_2014} on modern GPUs. Specifically, SBMM first launches a kernel that prepares the launch configuration, pointers to the weight, input and output addresses, and other necessary information for each delta, and then launches the actual blocked matmul kernel, which fuses dequantization for each delta and leverages sparse tensor cores on GPUs. The dynamic parallelism feature is revamped in the recent CUDA toolkit~\cite{cuda12releasenotes} and offers substantial performance improvements.
Figure~\ref{fig:kernel_launch} illustrates the kernel launch process for $3$ deltas and $4$ requests where the third delta has two requests. The first kernel is launched from the host and prepares the addresses of the weights, input and output for each delta, and then launches a blocked matmul for each delta in the second step. The actual matmul, in the last step, writes the results to the output. Figure~\ref{fig:kernel_launch_overhead} shows that even though the actual computation time is similar, our optimized kernel significantly reduces overhead and improves end-to-end latency.

We design our SBMM kernel to be compatible with popular low-precision and sparse matrix multiplication implementations. Optimizing such computations is an active research area for which many libraries have been developed, such as BitBLAS~\cite{wang_ladder_2024}, Marlin~\cite{frantar2024marlin}, and SparseMarlin~\cite{castro2024sparsemarlin}. Maintaining compatibility with these libraries allows us to leverage the latest hardware and community efforts. 

\subsection{Model Parallelism for Delta Serving}
\label{sec:model-parallelism}

\begin{figure}[tp]
  \centering
  \includegraphics[width=\linewidth]{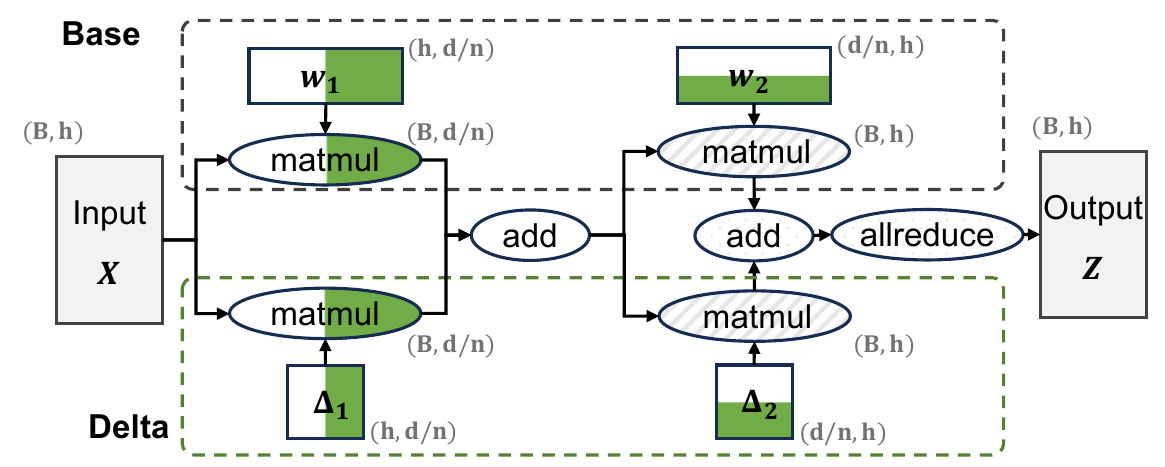}
  \caption{Tensor Parallelism in \projectname for $n$ = 2 GPUs. $B$ = number of tokens, $h$ = input dimension, $d$ = hidden size. Column-wise and row-wise partitions are illustrated as vertically and horizontally divided boxes, respectively.}
  \label{fig:model_parallelism}
\end{figure}

To serve large models that do not fit into a single GPU, \projectname extends Megatron-style~\cite{shoeybi_megatron-lm_2020} tensor parallelism to serve compressed deltas. In Megatron-style tensor parallelism, the model is partitioned column-wise or row-wise across multiple GPUs. \projectname adapts this to delta serving by  \textbf{partitioning the delta in the same way as the base model}. We first illustrate how our partition strategy works in Figure~\ref{fig:model_parallelism} with two linear layers, and then explain how we extend this approach to self-attention layers. Figure~\ref{fig:model_parallelism} assumes that we have two GPUs, a base model with two linear layers $[\mathbf{w}_1, \mathbf{w}_2]$ and a delta $[\Delta_1, \Delta_2]$. In the upper box, we partition the base model $\mathbf{w}_1$ to $[\mathbf{w}_{1,1}, \mathbf{w}_{1,2}]$ by column across two GPUs, and calculate the partial results $Y_{\text{base}, i}=X\mathbf{w}_{1,i}$ on each GPU. In the lower box, we partition the delta $\Delta_1$ to $[\Delta_{1,1}, \Delta_{1,2}]$ in the same way, and calculate the partial results $Y_{\text{delta}, i}=X\Delta_{1,i}$ on each GPU. The result of the matrix multiplication can be computed on each GPU independently without any synchronization with other GPUs as $Y=[Y_1, Y_2]=[Y_{\text{base}, 1} + Y_{\text{delta},1}, Y_{\text{base}, 2} + Y_{\text{delta},2}]$.

We then partition the second linear layer $\mathbf{w}_2, \Delta_2$ with row-parallel across two GPUs as $[\mathbf{w}_{2,1}, \mathbf{w}_{2,2}]^T, [\Delta_{2,1}, \Delta_{2,2}]^T$. Then the output of this layer becomes $Z=[Y_1,Y_2]\cdot [\mathbf{w}_{2,1}+\Delta_{2,1}, \mathbf{w}_{2,2}+\Delta_{2,2}]^T$. To compute this, we first perform $Y_i\cdot \mathbf{w}_{2,i}$ and $Y_i\cdot \Delta_{2,i}$ on each GPU individually, and then sum the results across GPUs with an all-reduce operation as the output.

In the self-attention module, we partition the $\mathbf{q}, \mathbf{k}$ and $\mathbf{v}$ projections as column-wise linear layers and the output projection $\mathbf{o}$ as row-wise linear layers. We then apply the same strategy as we described above to compute the output of the transformer block.

\subsection{Continuous Batching and Scheduling with Delta}
\label{sec:continuous-batching}

Last but not least, \projectname optimizes inference by extending continuous batching~\cite{yu_orca_2022}, a standard technique to improve GPU utilization for LLM serving.
\projectname implements a software scheduler per set of GPU workers that form a tensor parallel serving group. At every iteration, the scheduler picks $K$ requests to serve concurrently on a first-come-first-served basis. These $K$ requests belong to at most $N$ deltas that can fit concurrently in GPU memory. After selecting the $K$ requests and $N$ deltas, the scheduler searches the request queue for all requests that belong to the selected $N$ deltas. The intuition behind this design is to maximize batching by allowing requests to \textit{skip the line} if and only if they are for one of the $N$ selected deltas. The scheduler then delivers the batch of requests to the Model Runner, which starts to load the requested deltas, performs prompt processing if needed, and then batches request decoding for different models.

\begin{figure}[tp]
\includegraphics[width=\linewidth]{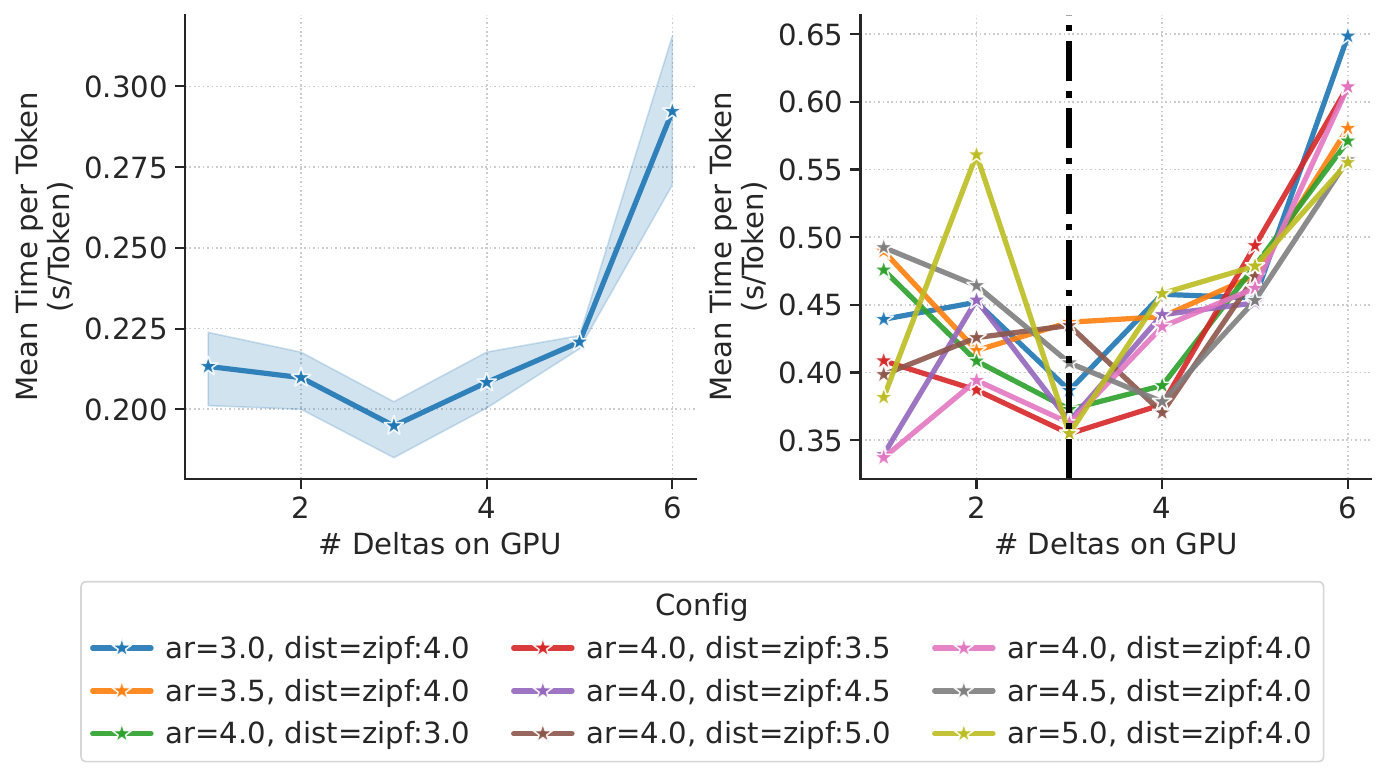}
    \caption{Normalized latency of \projectname with varying $N$, the number of concurrent deltas in GPU memory.}
\label{fig:offline-profiling}
\end{figure}

\textbf{Tuning $N$, the number of concurrent deltas.} \projectname tunes $N$ empirically based on a short trace profiling phase. We explore the impact of varying $N$. Intuitively, processing more deltas concurrently (larger $N$) allows for more extensive request batching, enhancing throughput. However, collocating many deltas increases GPU memory pressure, potentially degrading performance if there are already many requests to batch for a particular delta. We conduct a microbenchmark on RTX 3090 GPU and synthetic trace to study the effectiveness of offline profiling. On the left side of Figure~\ref{fig:offline-profiling}, we show how $N$ affects the serving latency when executing a 25 second time interval of a trace with arrival rate $3$ and zipf-4.0 model popularity distribution. In this case, \projectname would pick $N=3$ as it achieves optimal performance during profiling. We then run a series of traces under different settings. The right side of Figure~\ref{fig:offline-profiling} shows that $N=3$ remains optimal or near-optimal across different arrival rates and model popularity distributions. We generally find that offline profiling with a short trace is sufficient to determine the (near-)optimal number of concurrent deltas. Dynamic tuning can also be implemented. If offline profiling is not possible, operators can heuristically set $N$. If there are only few requests per delta, we can allow more deltas (and hence more requests in parallel) to improve the GPU utilization. If there are many requests per delta, we can reduce $N$ to avoid memory pressure. Like other systems, operators can also adjust the maximum batch size and concurrent deltas as hyperparameters.

\textbf{Avoiding starvation.} Allowing requests to skip the line is good for batching, but it may cause starvation if requests for the currently selected deltas keep arriving and skipping the line before the system has a chance to swap and serve other deltas. When a request skips the line to be batched with a currently loaded delta, we refer to the original request for that delta at the head of the queue as the \textit{parent request}. To alleviate starvation, \projectname uses a preemption strategy: \textbf{requests that skip the line are preempted when their parent request finishes}. Preempted requests get reinserted in their original place in the request queue (as if they did not skip the line) and can get scheduled in the next iteration. \projectname currently swaps the intermediate state of preempted requests to CPU memory and resumes computation when the request is re-scheduled. Future work involves exploring whether and when recomputing from scratch may be faster than swap-and-resume.

\noindent\textbf{Scalability.} \projectname employs a hierarchical management strategy when the number of deltas exceeds both GPU and host memory capacity. When there are more deltas than the capacity of GPU memory and even host memory, \projectname will swap deltas in and out of GPU/CPU memory to disk, and only load deltas when they are needed. For an inference batch, only a fixed and pre-set number of deltas (or less) will be processed concurrently and thus a single batch will not exceed the GPU memory capacity. The performance of \projectname will gracefully degrade if there are sporadic requests for deltas that reside on disk due to swapping.
\section{Evaluation}

\subsection{Performance Evaluation Setup}
\label{subsec_setup}

\textbf{Experiment testbed.} We conduct our experiments on a homogeneous GPU cluster. Each node has 2$\times$ Intel Xeon 8358P CPUs (128 threads) and 2TB DDR4 memory. We use 4 A800 GPUs per node and GPUs are interconnected to each other by NVLink and NVSwitch~\cite{nvlink}. Additionally, the cluster adopts an all-NVMe shared parallel file system, ensuring rapid data access and storage, connected through a 50Gbps RoCE network. All the experiments are conducted on this cluster unless explicitly stated.

\textbf{Models and downstream tasks.} We use the Llama-2~\cite{touvron_llama_2023} model with 7B, 13B, and 70B parameter and their fine-tuned variants. For 7B and 13B models we use the \texttt{Vicuna-v1.5}~\cite{vicuna2023} fine-tuned models since the fine-tuning data is disclosed and can be used for calibration. For 70B model we use the \texttt{Llama-2-70B-chat-hf} variant, and we use the fine-tuning dataset from \texttt{Vicuna}~\cite{vicuna2023} as a proxy to calibrate the compression. We mainly focus on serving 7B and 13B models in the serving experiments and compress them to 4-bit with 50\% sparsity. For these large models, we evaluate the post-compression quality on standard benchmarks in \textit{lm-eval-harness}~\cite{eval-harness}. We also fine-tune a smaller scale \texttt{Pythia-2.8B} model on \textit{natural instructions} tasks and evaluate the quality on the known downstream tasks. Since the accuracy drop is already substantial for SparseGPT and we found that the accuracy drop is more significant with lower precision, we do not evaluate the 2-bit 50\% sparsity for SparseGPT.

\begin{table}[!t]
    \centering
    \renewcommand{\arraystretch}{1.1}
    \resizebox{\linewidth}{!}{
  \begin{tabular}{@{}clcccr@{}}
    \toprule
  & \multicolumn{1}{c}{} & \multicolumn{3}{c}{\textbf{Downstream Tasks$\uparrow$}} &\\ \cmidrule(lr){3-5}
    \multirow{-2}{*}{\textbf{\begin{tabular}[c]{@{}c@{}}Base\\ Model\end{tabular}}} & \multicolumn{1}{c}{\multirow{-2}{*}{\textbf{Method}}} & T1  & T2 & T3 & \multirow{-2}{*}{\textbf{\begin{tabular}[c]{@{}c@{}}Compress\\ Ratio\end{tabular}}} \\ \midrule
  & FP16& 73.76& 94.23  & 79.61  & \cellcolor[HTML]{EFEFEF}1.00$\times$\\
  & SparseGPT (4bit$\star$)   & 67.59   & 91.99  & 72.67  & \cellcolor[HTML]{EFEFEF}3.93$\times$\\
    & \projectname (4bit$\star$)& 73.13   & \textbf{94.30} & \textbf{79.52} & \cellcolor[HTML]{EFEFEF}4.75$\times$\\
    \multirow{-4}{*}{\begin{tabular}[c]{@{}c@{}}Pythia\\ 2.8B\end{tabular}} & \projectname (2bit$\star$)& \textbf{74.44}  & 94.22  & 78.90  & \cellcolor[HTML]{EFEFEF}\textbf{8.36$\times$}   \\ \midrule
  & FP16  & 80.92 & 41.65  & 27.34  & \cellcolor[HTML]{EFEFEF}1.00$\times$\\
  & SparseGPT (4bit$\star$)& 67.16& 36.48  & 24.12  & \cellcolor[HTML]{EFEFEF}5.61$\times$\\
  & AWQ (4bit)& 81.22&  \textbf{42.24}& 27.34& \cellcolor[HTML]{EFEFEF}3.64$\times$\\
  & \projectname (4bit$\star$)& 81.41   & 41.72  & 27.50  & \cellcolor[HTML]{EFEFEF}5.39$\times$\\
    \multirow{-4}{*}{\begin{tabular}[c]{@{}c@{}}Llama\\ 7B\end{tabular}}& \projectname (2bit$\star$)& \textbf{81.56}& 41.91 & \textbf{28.26} & \cellcolor[HTML]{EFEFEF}\textbf{10.36$\times$}  \\ \midrule
  & FP16& 85.29& 43.00& 27.04& \cellcolor[HTML]{EFEFEF}1.00$\times$\\
  & SparseGPT (4bit$\star$)   & 79.88   & 35.01  & 23.20  & \cellcolor[HTML]{EFEFEF}5.91$\times$\\
  & AWQ (4bit)& 84.80& \textbf{43.37} & \textbf{27.80} & \cellcolor[HTML]{EFEFEF}3.82$\times$\\
  & \projectname (4bit$\star$)& \textbf{85.11}  & 42.48 & 27.04  & \cellcolor[HTML]{EFEFEF}5.91$\times$\\
    \multirow{-4}{*}{\begin{tabular}[c]{@{}c@{}}Llama\\ 13B\end{tabular}} & \projectname (2bit$\star$)& 84.95 & 42.54  & 27.65  & \cellcolor[HTML]{EFEFEF}\textbf{11.83$\times$}\\ \midrule
  & FP16 & 86.73 & 44.25 & 33.18 & \cellcolor[HTML]{EFEFEF}1.00$\times$\\
  & SparseGPT   (4bit$\star$) & 85.87 & 37.06  & 27.80 & \cellcolor[HTML]{EFEFEF}6.11$\times$\\
  & AWQ (4bit)& 86.36& 44.04& 31.80& \cellcolor[HTML]{EFEFEF}3.72$\times$\\
  & \projectname (4bit$\star$)& \textbf{87.28}& \textbf{44.18} & 32.87  & \cellcolor[HTML]{EFEFEF}5.84$\times$\\
    \multirow{-4}{*}{\begin{tabular}[c]{@{}c@{}}Llama\\ 70B\end{tabular}}   & \projectname (2bit$\star$)& 86.67& 43.47  & \textbf{33.49} & \cellcolor[HTML]{EFEFEF}\textbf{13.96$\times$} \\ \midrule
    
    & FP16& 83.70& 45.10& 27.65& \cellcolor[HTML]{EFEFEF}1.00$\times$\\
  & SparseGPT (4bit$\star$) & 74.86 & 38.79  & 21.50  & \cellcolor[HTML]{EFEFEF}1.53$\times$\\
  & AWQ (4bit) & 83.14 & \textbf{44.97} & \textbf{28.11}& \cellcolor[HTML]{EFEFEF}2.33$\times$\\
  & \projectname (4bit$\star$) & \textbf{83.79} & 44.24 & 27.50 & \cellcolor[HTML]{EFEFEF}3.26$\times$\\
    \multirow{-4}{*}{\begin{tabular}[c]{@{}c@{}}{Gemma 2}\\ 2B\end{tabular}}   & \projectname (2bit$\star$) & 83.66 & 44.40 & 26.26 & \cellcolor[HTML]{EFEFEF}\textbf{4.08$\times$} \\ \midrule
  
    & FP16& 88.87& 51.75 & 33.18& \cellcolor[HTML]{EFEFEF}1.00$\times$\\
  & SparseGPT   (4bit$\star$) & 86.26 & 45.71 & 29.95  & \cellcolor[HTML]{EFEFEF}2.43$\times$\\
  & AWQ (4bit) & 88.47 & 51.24& 33.33& \cellcolor[HTML]{EFEFEF}3.10$\times$\\
  & \projectname (4bit$\star$) & \textbf{88.96} & \textbf{51.32} & \textbf{34.87} & \cellcolor[HTML]{EFEFEF}4.61$\times$\\
    \multirow{-4}{*}{\begin{tabular}[c]{@{}c@{}}{Gemma 2}\\ 9B\end{tabular}} & \projectname (2bit$\star$) & 88.68 & 50.95 & 33.94 & \cellcolor[HTML]{EFEFEF}\textbf{7.50$\times$}
    \\ \bottomrule
  \end{tabular}}
     \caption{Model quality of \projectname vs. uncompressed (FP16) vs. SparseGPT~\cite{frantar_sparsegpt_2023} and AWQ~\cite{AWQ}. For \texttt{Pythia}, T1, T2, T3 show the accuracy on 3 downstream tasks (Amazon Review Classification, Synthetic Palindrome Numbers, Yes/No Question) from \textit{natural instructions}~\cite{mishra_cross-task_2022}. For other models, T1, T2, T3 show accuracy on 3 standard benchmarks (BoolQA, TruthfulQA, LogiQA) in \textit{lm-eval-harness}~\cite{eval-harness}. $\star$ indicates 50\% structured pruning in addition to quantization.}
    \label{tab:model_accuracy_large_simplified}
  \end{table}

\textbf{Workload traces.} To evaluate the serving performance, we use the prompts and responses sampled from the LMSys Chatbot Arena~\cite{zheng_judging_2023}. We consider three types of model popularity distribution: 1) Uniform: all models are equally popular. 2) Skewed: model popularity follows a Zipf-$\alpha$ distribution. In other words, the popularity of the $i$-th model is proportional to $1/i^\alpha$. We set $\alpha=1.5$ for the skewed distribution. 3) Azure trace: since there are no publicly available traces for multi-variant LLM serving available, following previous work~\cite{li_alpaserve_2023}, we use the Azure serverless function traces~\cite{shahrad_serverless_2020,zhang_faster_2021} as a proxy. Note the traffic in this trace is highly bursty and the model distribution is highly skewed. Except for the small-scale timeline analysis in Figure~\ref{fig:timeline_evaluation} (which we show for illustration purposes) and ablation studies, we run the traces for 5 minutes under different arrival rates and model distribution. Unless otherwise stated, there are 32 model variants that need to be served. The requests are sent to the serving system with a varying Poisson arrival rate (applied to the entire system).

\textbf{Metrics.} We use downstream accuracy to evaluate the post-compression quality and three metrics for serving performance: 1) end-to-end (E2E) latency averaged over all requests, 2) time to first token (TTFT) averaged over all requests, which is an important indicator of the system's responsiveness, 3) throughput and 4) SLO attainment (i.e., percentage of requests served within a given SLO requirement) for TTFT and E2E Latency.

\textbf{Baselines.}
For compression quality, we compare \projectname with SparseGPT~\cite{frantar_sparsegpt_2023} which incorporates both quantization and sparsification, as well as AWQ~\cite{AWQ} which is a state-of-the-art quantization algorithm. For serving performance, since there is no existing system that can serve multiple full-parameter fine-tuned models, we implement a simple baseline based on vLLM that supports 1) \underline{S}wapping models in and out of GPU memory, 2) \underline{C}ontinuous batching of different full-precision models by looping through models in a batch and 3) \underline{B}atching available requests for the same model. We refer to this baseline as \texttt{vLLM-SCB}. We use the maximum number of models that can fit in the GPU memory for the baseline serving system. We use a tensor parallelism degree of 4 for both \projectname and the baseline system unless otherwise stated. We do not show a comparison to ServerlessLLM~\cite{fu2024serverlessllm} as it treats each model as a black box --- meaning it does not batch requests from  models derived from the same base --- and  hence is not competitive for the scenarios we explore.

\textbf{System Setup.}
We evaluate our serving system without tuning the maximal number of concurrent deltas, since it requires offline profiling and apriori knowledge of the workload. Instead, we show two representative configurations to reveal the system performance under different settings without explicit tuning.

\subsection{Post-Compression Model Quality}
\label{subsec:quality}
We first study how inference accuracy is impacted by applying quantization and sparsification on model deltas compared to directly on FMT model weights.
As shown in Table~\ref{tab:model_accuracy_large_simplified}, \projectname achieves up to $13\times$ compression ratio with comparable accuracy to the original FP16 FMT models. On smaller models, \projectname achieves $4\times$ to $11\times$ compression ratio with comparable accuracy. We also found the compression ratio for Gemma 2 models is lower than the Llama models. This is because the Gemma 2 models have a relatively larger number of parameters stored in the embedding layers, which are not compressed.
In contrast, SparseGPT --- which applies similar techniques directly to the FMT model weights --- has a substantial drop in accuracy for all models. \projectname also achieves comparable accuracy compared with the state-of-the-art quantization algorithm AWQ~\cite{AWQ} but with a much higher compression ratio. We also observe that in some cases, the accuracy of the compressed model is even higher than the full-precision model. This effect is also noticeable in other works~\cite{han2015deep, AWQ, frantar_gptq_2023} and is likely due to the regularization effects of compression.
These results show that \algname can aggressively compress model deltas while retaining higher accuracy than the traditional approach of compressing model weights directly. In \S\ref{subsec:compare_to_lora}, we also show that compressed delta serving achieves higher accuracy on complex tasks than LoRA serving.

\begin{figure}[tp]
    \vspace{-10pt}
    \includegraphics[width=\linewidth]{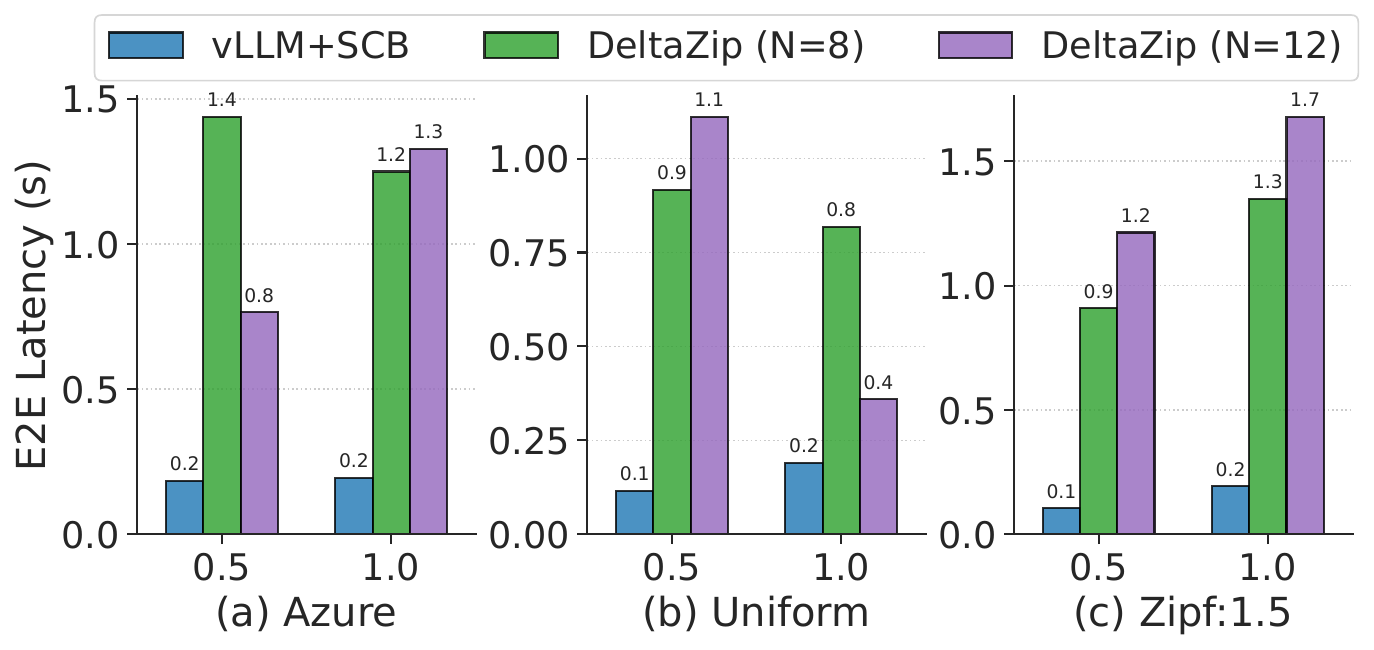}
    \caption{Throughput of different serving systems with varying poisson arrival rate $\lambda\in\{0.5, 1.0\}$ and distribution $\mathcal{D}\in\{\text{azure}, \text{uniform}, \text{zipf-1.5}\}$ for 13B model.}
    \label{fig:throughput}
\end{figure}

\begin{figure}[tp]
    \vspace{-10pt}
    \centering
    \begin{subfigure}{\linewidth}
        \includegraphics[width=\textwidth]{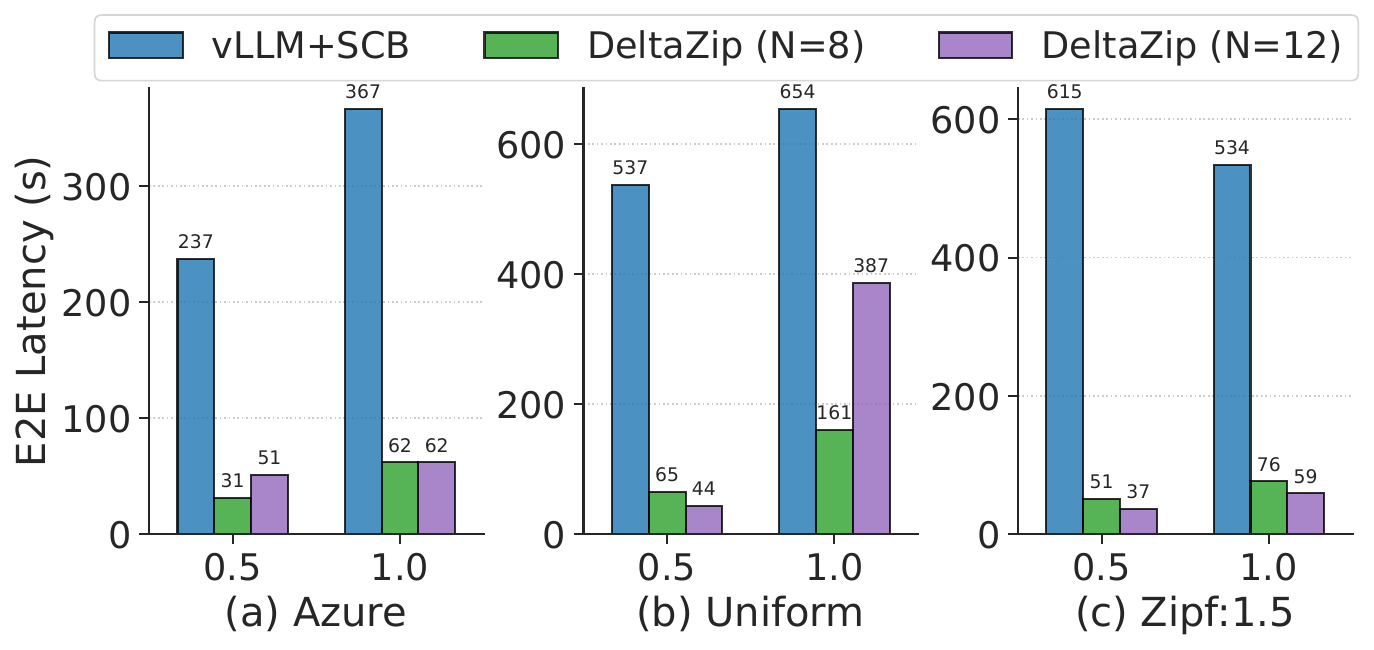}
        \caption*{Average E2E Latency}
        \label{fig:mean_e2e_latency}
    \end{subfigure}
    \hfill
    \begin{subfigure}{\linewidth}
        \includegraphics[width=\textwidth]{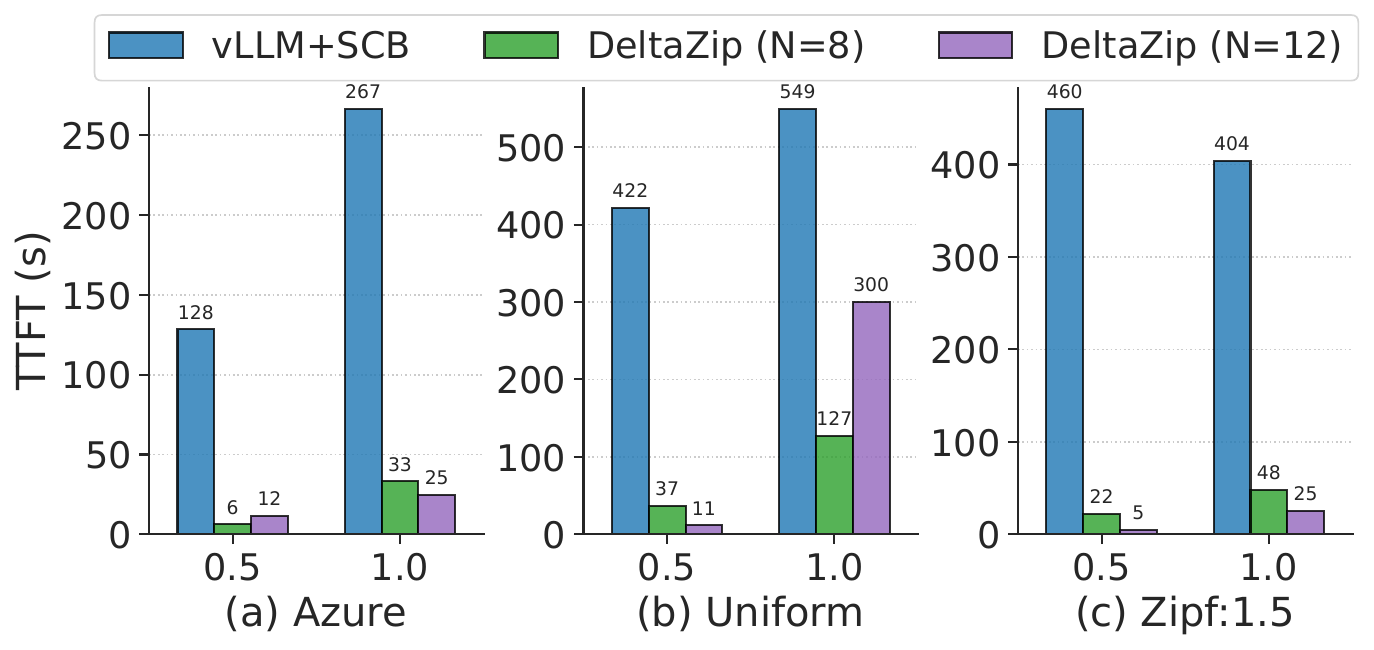}
        \caption*{Average TTFT}
        \label{fig:mean_ttft}
    \end{subfigure}
    \caption{Average latency of different serving systems with varying arrival rate and distribution for 13B model.}
    \label{fig:mean_latency}
\end{figure}

\subsection{End-to-end Serving Performance}
\label{subsec:serving_performance}

\textbf{Serving throughput.}
Figure~\ref{fig:throughput} shows the throughput of different serving systems with varying arrival rates and model distributions. We observe a 2$\times$ to 12$\times$ improvement in throughput compared to the baseline system. 
The improvement is more pronounced when the arrival rate is relatively light and more skewed. When the arrival rate is high and the model distribution is uniform, the improvement is less significant, and we find this is mainly due to the high cost in prompt processing when the models are more uniform. Since our techniques cannot reduce the prompt processing time, and when more deltas are batched together, requests in this batch have to wait for the prompt processing of the slowest request in the batch, leading to a lower throughput.

\begin{figure}[tp]
    \centering
    \begin{subfigure}{\linewidth}
        \includegraphics[width=\textwidth]{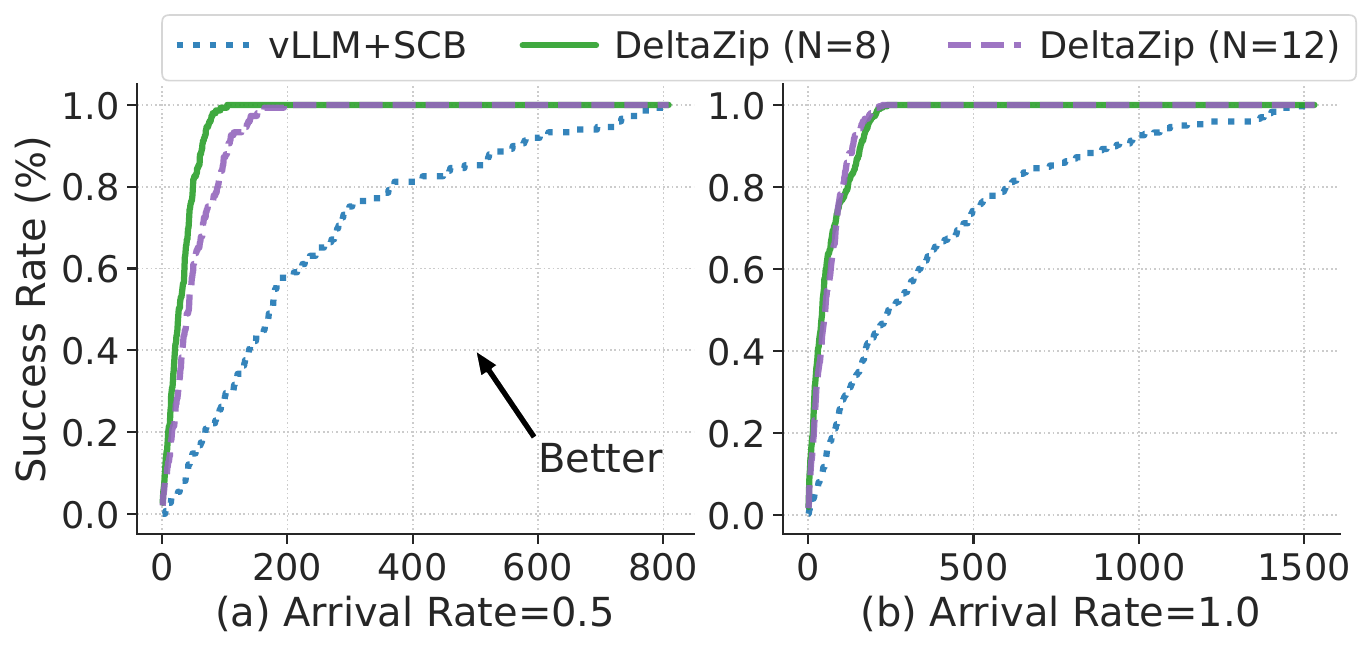}
        \caption*{SLO attainment of E2E latency}
        \label{fig:slo_e2e_latency}
    \end{subfigure}
    \hfill
    \begin{subfigure}{\linewidth}
        \includegraphics[width=\textwidth]{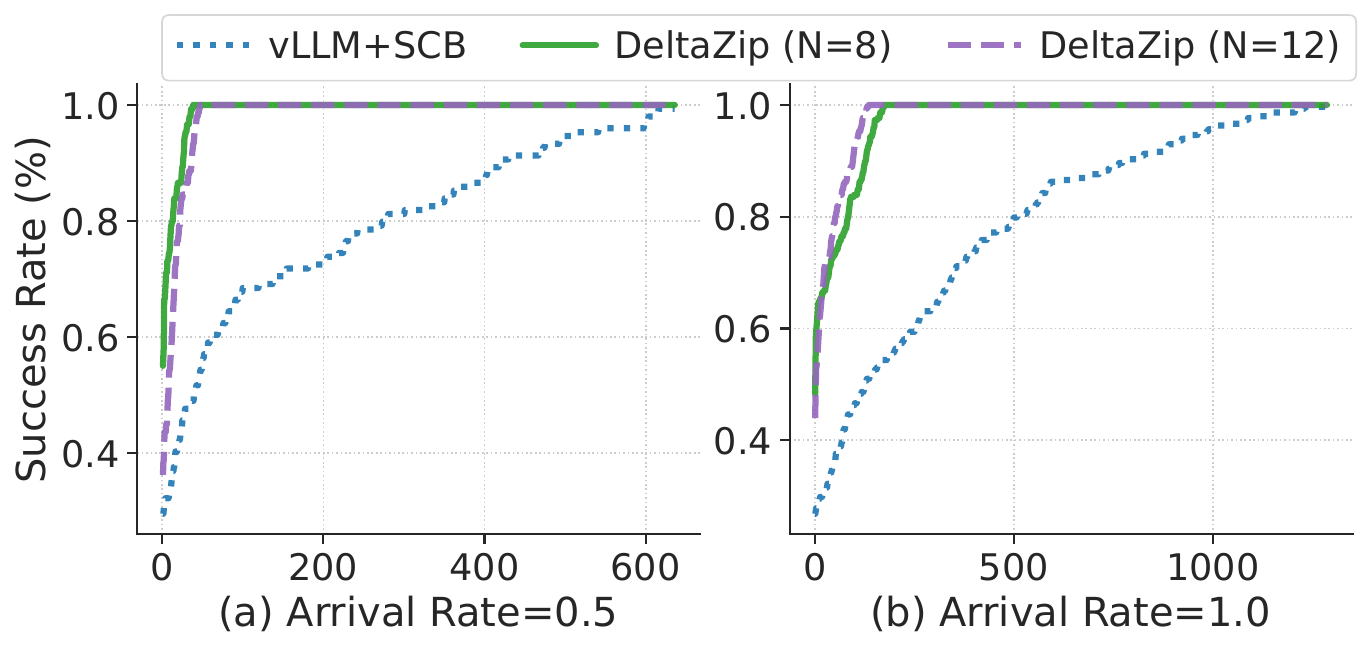}
        \caption*{SLO attainment of TTFT}
        \label{fig:slo_ttft}
    \end{subfigure}
    \caption{SLO attainment of different serving systems with varying arrival rate and azure model distribution for 13B model. X-axis denotes different SLO requirement in seconds.}
    \label{fig:slo_comparative}
\end{figure}

\textbf{Average E2E latency and TTFT.}
Figure~\ref{fig:mean_latency} shows that \projectname acheives a 1.6$\times$ to 16$\times$ improvement in average E2E latency and even higher improvement in TTFT. The high improvement in TTFT is due to \projectname's ability to batch more requests from different models concurrently, thus reducing queuing time. As in our throughput experiments, the improvement under uniform distribution with high load is less significant. We also observe that the maximum number of concurrent deltas being served has a substantial impact on the performance.

\textbf{SLO Attainment of TTFT and E2E Latency.}
Figure~\ref{fig:slo_comparative} shows the SLO attainment of E2E latency and TTFT with the Azure trace under varying Poisson arrival rate. We observe that \projectname achieves a higher SLO attainment compared to the baseline system.

\subsection{Delta versus LoRA Serving Approaches}
\label{subsec:compare_to_lora}

\begin{figure}[tp]
    \includegraphics[width=\linewidth]{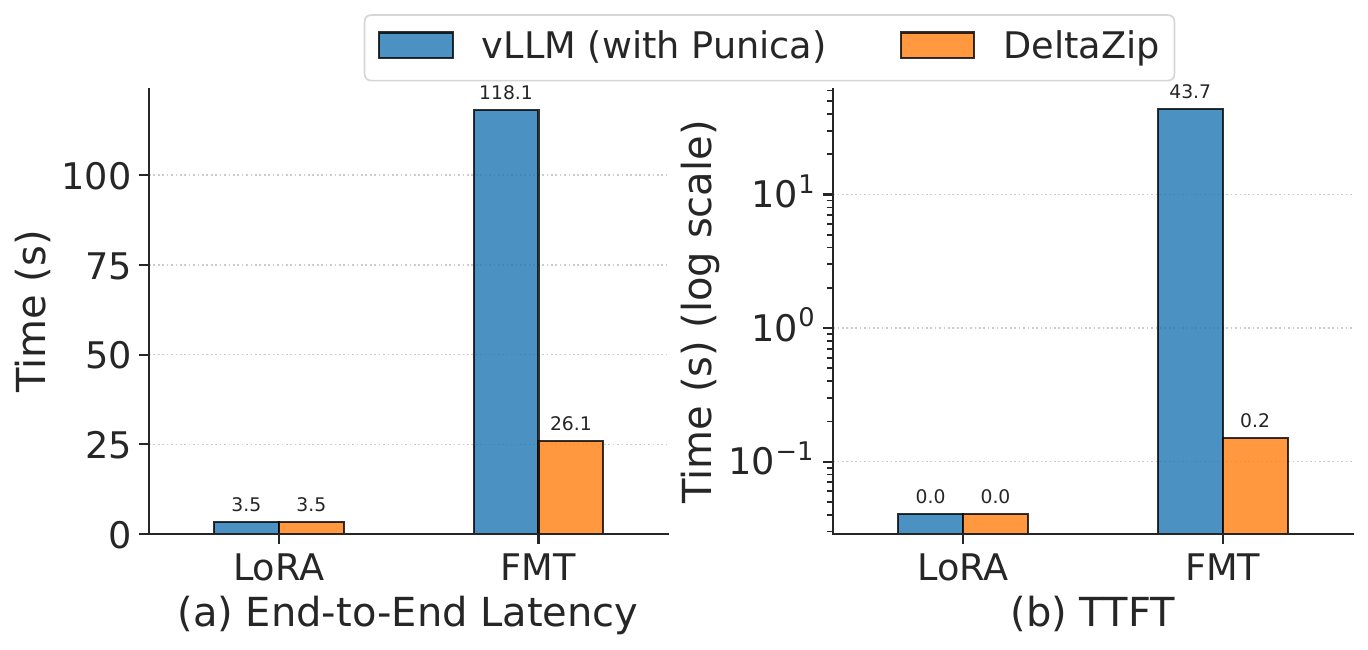}
    \caption{E2E latency and TTFT of \projectname and LoRA serving system when serving LoRA and FMT models.}
    \label{fig:compare_with_lora_2}
\end{figure}

\begin{table}[!t]
  \centering
  \renewcommand{\arraystretch}{1.1}
  \resizebox{\linewidth}{!}{
  \begin{tabular}{@{}clcccc@{}}
      \toprule
      \multirow{2}{*}{\textbf{Base Model}} & \multirow{2}{*}{\textbf{Task}} & \multicolumn{3}{c}{\textbf{Accuracy$\uparrow$}} \\
      \cmidrule(lr){3-5}
      & & FMT & LoRA & \algname \\
      \midrule
      \multirow{1}{*}{Llama-7B} & Math (GSM8K) & 34.79 & 29.49 & 34.95 \\
      \midrule
      \multirow{3}{*}{Pythia-2.8B} & Amazon Review & 73.76 & 50.92 & 73.13 \\
      & BoolQ Yes/No & 79.61 & 63.76 & 79.52 \\
      & NLI Classification & 73.23 & 52.15 & 70.74 \\
      \midrule
      \multirow{3}{*}{OpenLlama 3B} & Amazon Review & 76.07 & 55.07 & 77.36 \\
      & BoolQ Yes/No & 83.38 & 65.69 & 83.29 \\
      & NLI Classification & 80.07 & 63.46 & 79.75 \\
      \bottomrule
  \end{tabular}
  }
  \caption{Model quality of FMT vs. LoRA vs. \algname.}
  \label{tab:model_accuracy_vs_lora}
\end{table}

Complementary to LoRA serving systems, \projectname extends their serving
optimizations to also efficiently serve FMT models. \projectname inherits the
ability to serve LoRA adapters from vLLM~\cite{kwon_efficient_2023}, which
supports Punica-based~\cite{chen_punica_2023} LoRA serving.
Figure~\ref{fig:compare_with_lora_2} shows an example of using \projectname to
serve LoRA adapters on one GPU node and FMT model variants on another GPU node.
For LoRA serving, \projectname achieves similar performance as vLLM with Punica
kernels and for FMT serving, \projectname significantly outperforms the
baseline due to its delta compression approach.

Given that \projectname supports both LoRA and FMT model serving, a natural
question that arises for users is when to use which type of model fine-tuning
approach. While a detailed study is outside the scope of this paper and prior
works have already compared FMT and LoRA
accuracy~\cite{anyscale_fine-tuning_2023,lora-learns-forgets-biderman2024lora},
we conduct a brief analysis in which we apply delta compression for FMT
serving. Table~\ref{tab:model_accuracy_vs_lora} compares the accuracy of
uncompressed FMT models, LoRA adapters and \algname compressed FMT models. We
conduct extensive hyper-parameter tuning for LoRA adapters, using scripts from
Anyscale~\cite{anyscale_fine-tuning_2023}. We observe that even on complicated
tasks (e.g., Math) where LoRA adapters cannot achieve similar accuracy as FMT
models, \algname can maintaining high accuracy while compressing the FMT
models.

\begin{figure}[tp]
    \includegraphics[width=\linewidth]{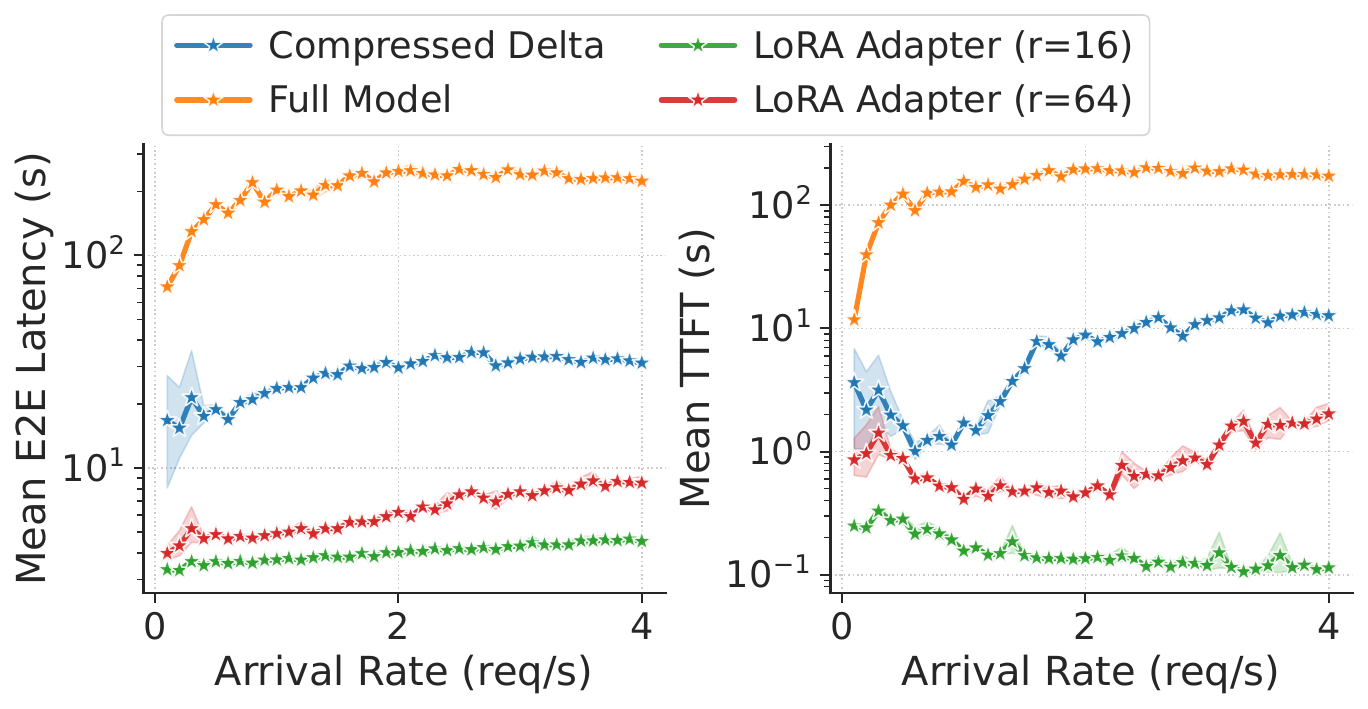}
    \caption{End-to-end latency and TTFT of \projectname and LoRA serving system with varying arrival rate.}
    \label{fig:compare_with_lora_1}
\end{figure}

We also compare the inference latency and TTFT of LoRA adapter, compressed
delta FMT and baseline full model FMT serving with varying arrival rates.
Figure~\ref{fig:compare_with_lora_1} shows that compressed deltas and LoRA
adapters are much more efficient to serve than the vLLM+SCB baseline approach
for FMT model serving, which swaps full models. Serving LoRA adapters is still
more efficient than serving compressed deltas, mainly due to the more compact
size and smaller memory footprint of LoRA adapters. In conclusion,
\textbf{users should choose between LoRA and compressed delta FMT serving based
    on the trade-off between accuracy and serving performance}: LoRA is more
suitable for tasks where accuracy is not the primary concern or for simpler
tasks where LoRA can achieve comparable accuracy to full model tuning. In
contrast, compressed delta FMT serving is more suitable for tasks where
accuracy is critical. \projectname improves the serving efficiency for FMT
serving with the \algname delta compression approach and its optimized delta
serving system design and implementation.

\subsection{Microbenchmarks and Ablation Study}
\label{sec:microbenchmarks}
\begin{figure}[tp]
    \centering
    \includegraphics[width=\linewidth]{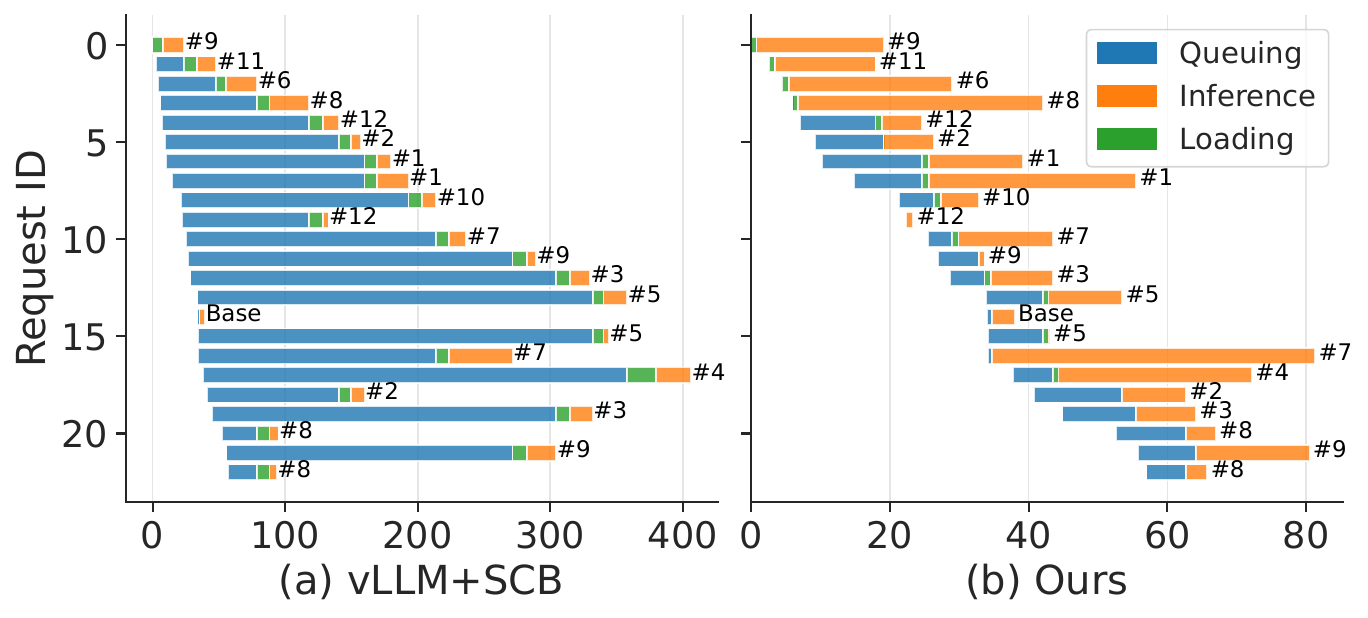}
    \caption{Serving latency breakdown. The \# on the right indicates the model ID. The x-axis is the time in seconds.}
    \label{fig:timeline_evaluation}
\end{figure}

\textbf{Latency breakdown.} We perform a smaller-scale experiment to visualize the latency breakdown of different serving stages (queuing, loading, and inference). We synthesize a trace with 12 models, arrival rate of $0.5$ requests per second for 60 seconds and run on two RTX 3090 GPUs. Figure~\ref{fig:timeline_evaluation} compares \projectname with the baseline. The vLLM+SCB system has two main issues: 1) the loading time is substantial, as it needs to load the entire model from the disk to GPU, and 2) queuing time dominates, due to the lack of batching. Even though it batches requests for the same model (e.g., for the model \#2), it cannot batch requests from many different models due to GPU memory capacity limits. In contrast, \projectname alleviates these two issues by 1) loading only the compressed deltas, which can be $5\times$ to $10\times$ smaller than the full model, and 2) batching requests from different models, which significantly reduces queuing delays.

\begin{figure}[tp]
    \centering
    \includegraphics[width=\linewidth]{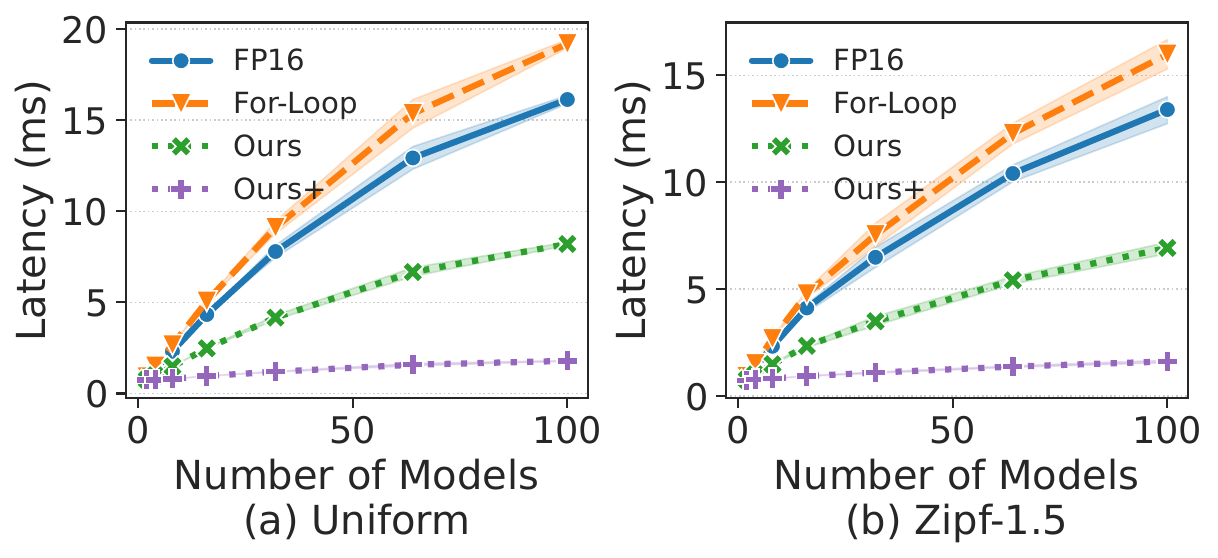}
    \caption{Microbenchmark of the SBMM kernel vs. the baseline implementation on a single GPU with varying number of models given a fixed number of requests. ``Ours'' refers to the implementation with reduced random memory access only and ``Ours+'' refers to the kernel implementation we proposed in \S\ref{sec:base-and-delta-decoupling}. For FP16, we do not perform the decoupling as it will only introduce additional overhead.}
    \label{fig:sbmm_kernel_micro_benchmark}
\end{figure}

\textbf{SBMM kernel.} We evaluate the performance of the SBMM kernel described in \S\ref{sec:base-and-delta-decoupling}. Figure~\ref{fig:sbmm_kernel_micro_benchmark} shows the performance of the SBMM kernel compared to the baseline implementations as we scale the number of models. We observe that the naive for-loop approach does not bring much performance improvement and is comparable with the half-precision implementation. However, with reduced random memory access, we observe a $2\times$ speedup compared to the baseline implementation, and the proposed kernel further improves the performance by another $2\times$ to $3\times$. In addition, we also observe that given the same number of requests, our kernel scales well with different number of models.

\begin{figure}[tp]
    \includegraphics[width=\linewidth]{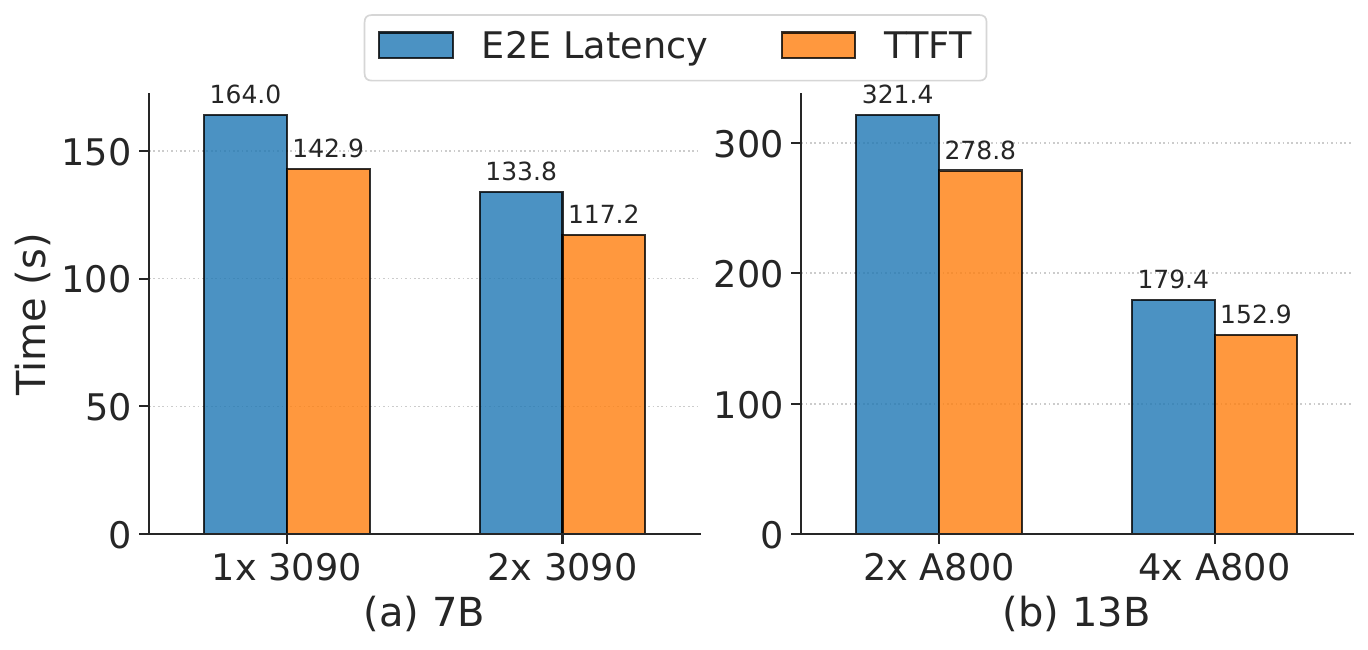}
    \caption{End-to-end latency and time to first token (TTFT) of \projectname with varying number of GPUs.}
    \label{fig:tp_experiment}
\end{figure}

\textbf{Model Parallelism.} We also conduct an experiment with varying numbers of GPUs under different settings to evaluate the model parallelism of \projectname. Figure~\ref{fig:tp_experiment} shows the end-to-end latency and TTFT of \projectname with varying number of GPUs. We observe that the latency decreases with the number of GPUs, particularly on A800 platform. This is because the inter-GPU communication is faster on A800 platform compared to the RTX 3090 platform. This observation leaves a direction for future work to optimize and tune the tensor parallelism degree of \projectname.

\begin{figure}[tp]
    \vspace{-10pt}
    \includegraphics[width=\linewidth]{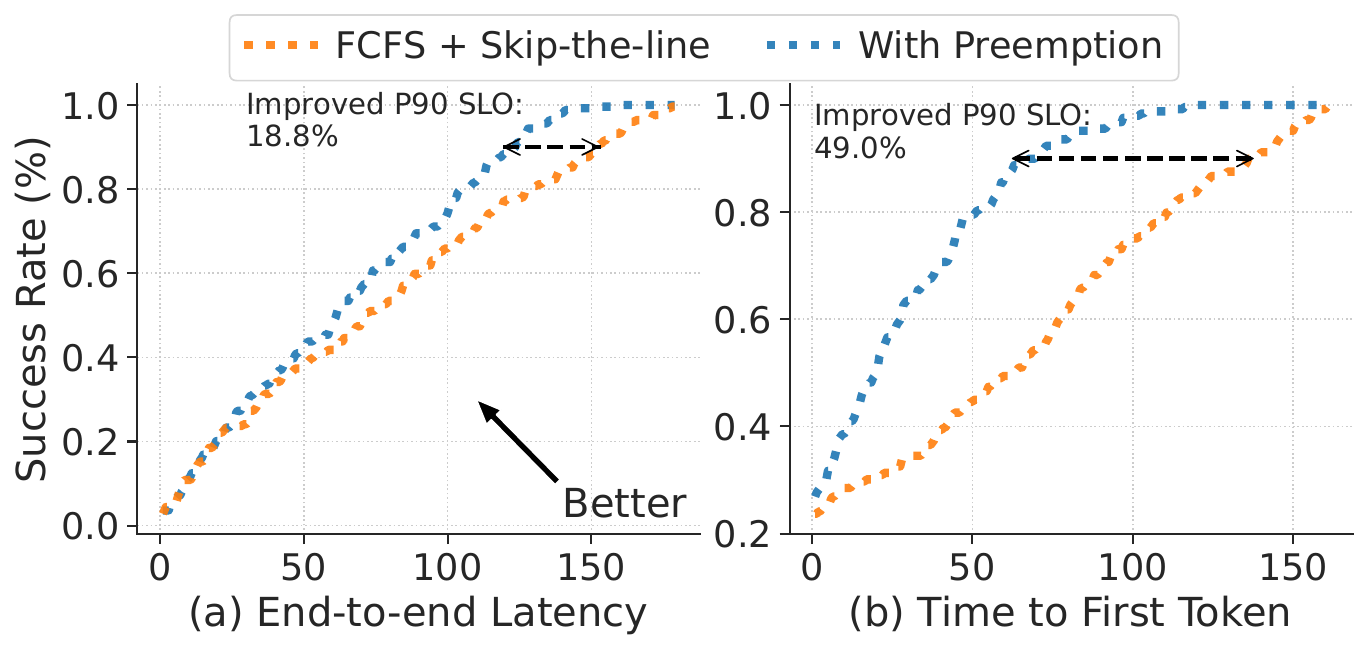}
    \caption{E2E Latency and TTFT of \projectname with and without preemption.}
    \label{fig:pre-emption}
\end{figure}

\textbf{Starvation Handling.} Next we evaluate the effectiveness of the preemption mechanism as described in \S\ref{sec:continuous-batching}. Figure~\ref{fig:pre-emption} shows the E2E latency and TTFT with and without preemption. We observe that the preemption mechanism effectively reduces the latency, particularly the time-to-first-token as it allows more requests to start earlier.

\section{Related Work}
\label{sec:related_work}

\textbf{ML Model Serving Systems.}
Optimizing ML serving is an active area of research~\cite{gujarati_serving_2020,romero_infaas_2021,crankshaw_clipper_2017}. With the increasing popularity of LLMs, there has been a surge of LLM serving systems proposing optimizations such as GPU kernel implementations~\cite{dao_flashattention_2022,dao_flashattention-2_2023,xFormers2022,flashinfer}, advanced and fine-grained batching~\cite{fang_turbotransformers_2021, yu_orca_2022}, memory management~\cite{kwon_efficient_2023}, and parallelism~\cite{li_alpaserve_2023,shoeybi_megatron-lm_2020}. However, most of these works do not consider the multi-variant serving scenario, for which \projectname is designed. For multi-model LLM serving, MuxServe~\cite{duan_muxserve_2024} explores spatial-temporal multiplexing for LLM serving to improve GPU utilization. ServerlessLLM~\cite{fu2024serverlessllm} studies the feasibility of serving LLMs in a serverless environment and proposes fast checkpoint loading and locality-driven live migration. However, these works treat the models as a black box and do not consider their lineage, and are hence orthogonal to \projectname.
Punica and S-LoRA~\cite{chen2024punica,sheng_s-lora_2023} are the most relevant to \projectname, as they also target multi-variant model serving scenarios, but they focus on serving LoRA adapters and do not consider serving full fine-tuned models. \projectname is complementary to these works.

\textbf{Post-Training Compression.}
Another line of work focuses on reducing the memory footprint of large models by compressing the model weights in a lossy manner. Beyond the OBS framework~\cite{hassibi_optimal_1993} and its improvements, such as GPTQ~\cite{frantar_gptq_2023} and SparseGPT~\cite{frantar_sparsegpt_2023}, many compression algorithms~\cite{AWQ,xiao_smoothquant_2023,yao_zeroquant_2022,tseng_quip_2024} have been proposed to reduce the model size. Such techniques can also be applied to model deltas by adapting the \projectname compression pipeline. There is also concurrent work on model delta compression. The initial version of our system~\cite{yao2023deltazip} and GPT-Zip~\cite{isik_gpt-zip_2023} proposes quantization and unstructured sparsity, but as this paper shows, with structured sparsity, we can achieve similar compression ratio and further improve the system performance. BitDelta~\cite{liu2024bitdelta} proposes extreme quantization and DARE~\cite{yu2024language} proposes unstructured sparsity. Compared to these works, \projectname optimizes more for \textbf{hardware efficiency} by combining quantization with \textit{structured sparsity}. To the best of our knowledge, \projectname is also the first serving system to support both LoRA and compressed delta FMT model serving.
\section{Discussion}
\label{sec:discussion}

\textbf{Limitations.}
While \projectname's decoupled computation improves throughput and reduces latency when concurrently serving many models, decoupled inference still has higher unloaded  latency than serving a FMT model directly in GPU memory. Hence, when there are only a handful of models to serve and they fit in GPU memory, \projectname may not be suitable. Compared to dedicated instances for each model, \projectname may be less performant, but it is more cost-effective and are more suited for environments where the cost and performance needs to be balanced. In addition, the co-serving of LoRA and FMT models in \projectname is at a coarse granularity, where LoRA and FMT models are served in two separate sets of GPUs and must be in separate batches. We plan to explore the possibility of serving LoRA and FMT models in the same batch as future work.
Furthermore, \projectname reorders requests to to group requests belonging to the same delta together, thus it cannot guarantee the SLO constraints of individual models. Potential future work includes adding mechanisms to prioritize models based on their constraints, as well as predicting the output length of each request to better guarantee SLOs. Finally, the starvation handling mechanism in \projectname is simple and may not be beneficial for all cases. For example, the preemption of requests that are about to finish leads to unnecessary starvation and performance degradation. We plan to explore more sophisticated mechanisms, such as output length prediction, to handle starvation in future work.

\textbf{Supporting PEFT approaches beyond LoRA.}
\projectname's decoupled computation architecture is general and can be used beyond LoRA, for other PEFT methods being proposed to improve accuracy.  For example, GaLore~\cite{zhao2024galore} only uses low-rank on the gradient but results in full-rank weight updates. RoSA~\cite{nikdan_rosa_2024} introduces sparse adapters in addition to low-rank adapters. Due to the lack of support for full-rank weight updates, these methods cannot be served by existing LoRA-based systems. We plan to extend \projectname to add support for emerging PEFT methods. 
\section{Conclusion}
\label{sec:conclusion}
To conclude, we propose \projectname, a serving system that enables LLM service providers to efficiently serve multiple fine-tuned models, whether they are fine-tuned through parameter-efficient or full-model tuning techniques.
For efficient serving of full-model-tuned models, \projectname leverages a key insight: fine-tuning typically results in small perturbations, allowing model deltas to be highly compressible. \projectname co-designs the serving system with the compression algorithm and achieves 10$\times$ compression ratio, improves serving throughput by 2$\times$ to 12$\times$ and maintains high model quality comparable to FP16 models. One practical use case of \projectname is to pack less-popular models on a limited pool of GPUs, and for those popular models, service providers can still deploy them on dedicated machines, which allows for better resource allocation and minimizes operational costs.

\begin{acks}
We would like to thank Berivan Isik, Hermann Kumbong, Wanyi Ning, Ce Zhang, Sanmi Koyejo for their contribution and feedback during the early stage of the project. We also thank Lijie Xu, Yongjun He, Wenqi Jiang, Foteini Strati, anonymous EuroSys reviewers, and our shepherd, Thaleia Dimitra Doudali, for their insightful reviews and feedback. This work is partially supported by a Google Academic Research Award.
\end{acks}

\appendix
\section{Artifact Appendix} 

\subsection{Abstract}

\subsection{Description \& Requirements}

Our open-source code for artifact evaluation can be found at: \url{https://github.com/eth-easl/deltazip-ae} as well as \url{https://doi.org/10.5281/zenodo.14870277}. We also provide a docker image (\texttt{ghcr.io/xiaozheyao/deltazip}) for easy setup of the software environment. The README file in the repository provides instructions on how to use the code and produce results.

\subsubsection{How to access}

Our code is available at: \url{https://github.com/eth-easl/deltazip-ae}. The docker image is available at: \texttt{ghcr.io/xiaozheyao/deltazip}.

\subsubsection{Hardware dependencies}

For the compression experiments, our reproduce script requires a machine with NVIDIA ampere GPUs (e.g., A100, RTX 3090, etc.) with equal or more than 24GB GPU memory. The machine should have at least 100 GB of storage to store the model weights and compressed models.

For the performance experiments, our reproduce script requires a machine with 4 NVIDIA A100 GPUs, and over 1TB of storage. The machine should have at least 480GB of RAM.

\subsubsection{Software dependencies}

We provide a docker image for easy setup of the software environment. In order to run the docker image, the host machine should have \texttt{Docker} and \texttt{nvidia-container-toolkit} installed. The host machine should also have installed \texttt{CUDA 12.0} and above.

If you prefer to run the code without the docker image, the software dependencies can be found in the \texttt{requirements.txt} file under \texttt{compression} and \texttt{serving} directories. We recommend refer to the \texttt{Dockerfile} for the exact dependencies of our artifact.

Notably, we also rely on \href{https://github.com/EleutherAI/lm-evaluation-harness}{lm-eval-harness} to evaluate the accuracy of the models.

\subsubsection{Benchmarks} 

\subsection{Set-up}

Please set up the environment variables and paths using \texttt{scripts/env.sh}.

\subsection{Evaluation workflow}

Our evaluation workflow consists of two parts: compression/model quality evaluation and performance evaluation.

\subsubsection{Major Claims}

Our major claims are that the model deltas can be compressed by a factor of \texttt{10x} into 2-bit per parameter and 50\% structured sparsity, with minimal loss in model quality. Such compression can be leveraged to improve the performance of the model serving system.

\subsubsection{Experiments}

\textbf{Experiment (E1)}: [Compressing 7B Model] [30 human-minutes + 3 compute-hours]: We compress the 7B model to 2-bit per parameter and 50\% structured sparsity. We expect the compressed model to have a compression ratio of 10x with minimal loss in model quality.

\begin{verbatim}
    bash scripts/1_compress_7b.sh
\end{verbatim}

The above command should take around 30 minutes to run.

We then test that the model can generate reasonable text samples.

\begin{verbatim}
    bash scripts/1.1_test_generate.sh
\end{verbatim}

We then evalaute the model quality using the \href{https://github.com/EleutherAI/lm-evaluation-harness}{lm-eval-harness} toolkit, by running the following command:

\begin{verbatim}
    bash scripts/2_eval_quality_compressed.sh
\end{verbatim}

The above commands will print the accuracy of the model in a table format.

To compare the model quality of the compressed model with the original model, we evaluate the performance of the original model using the following command:

\begin{verbatim}
    bash scripts/2_eval_quality_original.sh
\end{verbatim}

To compare the model quality with other baseline compression techniques, we first compress the models using SparseGPT and AWQ, by running the following commands:

\begin{verbatim}
    bash scripts/2.1_compress_baselines.sh
\end{verbatim}

Then we can evaluate the model quality of the compressed models using the following command:

\begin{verbatim}
    bash scripts/2.2_eval_baselines.sh
\end{verbatim}

We provide a helper script to summarize the results:

\begin{lstlisting}[breaklines,style=code]
    python scripts/helpers/aggregate_accuracy.py --compressed-model $WORKDIR/compressed_models/lmsys.vicuna-7b-v1.5.2b_2n4m_128bs --full-model lmsys/vicuna-7b-v1.5 --accuracy-dir $WORKDIR/eval_results --sparsegpt-model $WORKDIR/sparsegpt_models/lmsys.vicuna-7b-v1.5.4b_2n4m_128bs --awq-model $WORKDIR/awq_models/awq.lmsys.vicuna-7b-v1.5.4b128g
  \end{lstlisting}

\textbf{Experiment (E2)}: [Performance Evaluation] [60 human-minutes + 5 compute-hours]:

In order to evaluate the serving performance of our system, we first need to compress the 13b model, prepare the delta weights and create multiple copies.

\begin{verbatim}
    bash scripts/3_compress_13b.sh
    bash scripts/3_prepare_delta.sh
\end{verbatim}

We pre-partition the model weights into the tensor-parallel shards. If you change the tensor-parallel degree, you need to re-partition the model weights. By default, we set tensor parallel degree to be $4$.

In another terminal, run the following command to start the client:

\begin{lstlisting}[breaklines,style=code]
    python3 scripts/helpers/bench.py --workload scripts/workload/azure.ar=0.5.jsonl --base-model meta-llama/Llama-2-13b-hf --output $WORKDIR/results
  \end{lstlisting}

Note that the performance numbers may vary depending on the exact machine you are running the evaluation on. In general, we expect DeltaZip to have lower latency and higher throughput compared to the baseline.

\balance

\bibliographystyle{plain}
\bibliography{references.bib, others.bib}


\end{document}